\documentstyle[12pt,aasms4]{article}

\def \ang          {\AA}
\def \ref {\noindent\hangindent=1.0in\hangafter=1}
\def \cl {\centerline}

\def\ltsima{$\; \buildrel < \over \sim \;$}
\def\simlt{\lower.5ex\hbox{\ltsima}} 
\def\gtsima{$\; \buildrel > \over \sim \;$}
\def\simgt{\lower.5ex\hbox{\gtsima}} 

\begin{document}

\title{Multiwavelength Monitoring of the BL Lacertae Object PKS~2155-304
in May 1994. II. The IUE Campaign }


\author{Elena Pian\altaffilmark{1,2}, C. Megan Urry\altaffilmark{1,2}, 
Aldo Treves\altaffilmark{3,4}, Laura Maraschi\altaffilmark{5,2},}

\author{Steve Penton\altaffilmark{6}, J. Michael Shull\altaffilmark{6}, 
Joseph E. Pesce\altaffilmark{1,2}, Paola Grandi\altaffilmark{7}, }

\author{Tsuneo Kii\altaffilmark{8}, Ron I. Kollgaard\altaffilmark{9},
Greg Madejski\altaffilmark{10}, Herman Marshall\altaffilmark{11}, }

\author{Willem Wamsteker\altaffilmark{12}, Annalisa 
Celotti\altaffilmark{4,13}, Thierry J.-L. Courvoisier\altaffilmark{14},}

\author{Renato Falomo\altaffilmark{15}, Henner H. Fink\altaffilmark{16}, 
Ian M. George\altaffilmark{10}, Gabriele Ghisellini\altaffilmark{17}}

\altaffiltext{1}{Space Telescope Science Institute, 3700 San Martin Drive, 
Baltimore, MD 21218}
\altaffiltext{2}{Guest Observer with the International Ultraviolet Explorer}
\altaffiltext{3}{Dpt. of Physics, University of Milan, via Celoria 16, I-20133
Milan, Italy}
\altaffiltext{4}{Scuola Internazionale Superiore di Studi Avanzati, via Beirut
2-4, I-34014 Trieste, Italy}
\altaffiltext{5}{Osservatorio Astronomico di Brera, via Brera 28, I-20121 
Milano, Italy}
\altaffiltext{6}{University of Colorado, JILA, Campus Box 440, Boulder, 
CO 80309-0440}
\altaffiltext{7}{IAS/CNR, via Enrico Fermi 23, CP67, I-00044 Frascati, Italy}
\altaffiltext{8}{Institute for Space and Astronautical Science, 3-1-1 
Yoshinodai, Sagamihara, Kanagawa 229, Japan}
\altaffiltext{9}{Department of Astronomy and Astrophysics, The Pennsylvania
State University, University Park, PA 16802}
\altaffiltext{10}{Laboratory for High Energy Astrophysics, Goddard Space 
Flight Center, Greenbelt, MD 20771}
\altaffiltext{11}{Eureka Scientific, Inc., 2452 Delmer St., Suite 100,
Oakland, CA 94602}
\altaffiltext{12}{ESA IUE Observatory, P.O. Box 50727, 28080 Madrid, Spain}
\altaffiltext{13}{Institute of Astronomy, Madingley Road, Cambridge, 
CB3 0HA, United Kingdom}
\altaffiltext{14}{INTEGRAL Science Data Centre, 16 Chemin d'Ecogia,
CH-1290 Versoix, Switzerland}
\altaffiltext{15}{Osservatorio Astronomico di Padova, via dell'Osservatorio 5,
I-35122 Padova, Italy}
\altaffiltext{16}{Max Planck-Institute f\"ur Extraterrestrische Physik,
Giessenbachstrasse, 85740 Garching bei M\"unchen, Germany}
\altaffiltext{17}{Osservatorio Astronomico di Brera, via Bianchi 46,
I-22055 Merate, Italy}

\begin{abstract}

PKS~2155-304, the brightest BL Lac object in the ultraviolet sky, was monitored
with the IUE satellite at $\sim$1 hour time-resolution for ten nearly 
uninterrupted days in May 1994. The campaign, which was coordinated with
EUVE, ROSAT, and ASCA monitoring, along with optical and radio observations 
from the ground, yielded the largest set of spectra and the richest short 
time scale variability information ever gathered for a blazar at UV 
wavelengths. The source flared dramatically during the first day, with an 
increase by a factor $\sim$2.2 in an hour and a half. In subsequent days, the 
flux maintained a nearly constant level for $\sim$5 days, then flared with 
$\sim$35\% amplitude for two days. The same variability was seen in both
short- and long-wavelength IUE light curves, with zero formal lag ($\simlt$ 2 
hr), except during the rapid initial flare, when the variations were not 
resolved. Spectral index variations were small and not clearly correlated with
flux. The flux variability observed in the present monitoring is so rapid that
for the first time, based on the UV emission alone, the traditional 
$\Delta L /\Delta t$ limit 
indicating relativistic beaming is exceeded. The most rapid variations, under
the likely assumption of synchrotron radiation, lead to a lower limit of 1 
G on the magnetic field strength in the UV emitting region. These results are 
compared with earlier intensive monitoring of PKS~2155--304 with IUE in 
November 1991, when the UV flux variations had completely different 
characteristics. 
\end{abstract}



\section{Introduction}

Variability of active galactic nuclei (AGN) provides the clearest evidence for
dynamic processes occurring in the central engines and in the jets of these
objects. Its study is therefore a powerful way to investigate the innermost 
regions of AGN and the emission mechanisms responsible for the huge observed
luminosities.

The emission from blazars spans the range from radio to $\gamma$-ray energies,
and exhibits more rapid and higher amplitude variability than other AGN 
(Bregman 1990; Wagner \& Witzel 1995). Therefore, simultaneous multiwavelength
monitoring of blazars is particularly suited to estimating the sizes of the 
emitting regions (as a function of wavelength) and to understanding, through 
correlated variability at different frequencies, the radiation processes.

The most widely accepted picture for blazar emission at radio through UV
wavelengths is the synchrotron process within an inhomogeneous jet. The model 
is typically characterized by a spatial dependence of the magnetic field, 
electron density and maximum electron energy, and usually incorporates a 
relativistic velocity of the plasma within the jet, which causes beaming of 
the radiation. How the power is transported along the jet and transferred to 
the high energy electrons responsible for the observed emission is still 
unknown. Particle acceleration may take place at a single (dominant) shock 
front or in a quasi-continuous way (small shocks) along the jet. In the former
case, the spectral energy distribution from the millimeter to the soft X-rays 
derives from the energy distribution of the relativistic electrons accelerated
at the shock front, with lower energy particles extending farther from the
shock due to their longer lifetimes. In the case of {\it in situ} 
acceleration (Marscher 1980; K\"onigl 1981; Ghisellini, Maraschi, \& Treves 
1985; Hutter \& Mufson 1986) the maximum emitted synchrotron frequency usually
decreases along the jet, with UV and soft X-rays being produced closest to the
central active source.

In PKS~2155--304, the brightest known BL Lac object at UV wavelengths, 
synchrotron emission produces the optical and UV continuum, as demonstrated by
simultaneous spectropolarimetric observations in the two bands (Allen et al. 
1993). The synchrotron emission extends to the medium X-ray range (Kii et al. 
1996) and has a maximum power per decade ($\nu F_\nu$) between the UV and soft
X-ray range (Wandel \& Urry 1991). The spectral steepening from optical to UV
to X-rays can be attributed to radiative energy losses in the single shock 
model, or to the decreasing volume of the region emitting at higher 
frequencies in the inhomogeneous jet model. In either case the highest 
amplitude synchrotron variability is expected to be observed at or above the 
peak power output, which is determined by the steady-state balance of electron
acceleration and radiation, since small changes in the electron acceleration 
substantially alter the higher energy emission. 

Previous monitoring of PKS~2155--304 with IUE probed its variability in the 
far-UV domain (1200-3000 \AA) on a range of time scales from years down to a 
few hours, though the sampling was usually sparse, uneven, or limited in time 
(Maraschi et al. 1986; Urry et al. 1988; Treves et al. 1989; Edelson et al. 
1991; Urry et al. 1993, henceforth U93). The IUE campaign in November 1991 
(U93), which was coordinated with ROSAT observations, had adequate time 
coverage (30 days) and sampling to probe interday variability on an extended 
time interval, and even intraday variability during the continuous observing 
period ($\sim$5 days out of 30). The presence of recurrent flares on a 
$\sim$0.7-day time scale prompted further IUE intensive monitoring in May 
1994, coordinated with EUVE (Marshall et al. 1996), ASCA (Kii et al. 1996), 
and ROSAT (Urry et al. 1996), as well as radio, near-IR, and optical coverage
from ground-based telescopes (Pesce et al. 1996). The aim of the IUE campaign 
was to obtain the longest and best sampled UV light curve ever, in order to 
test the shortest possible variation time scales, within the capabilities of 
the IUE instruments, and to explore the correlation with emission at other 
wavelengths (Urry et al. 1996).

In this paper we concentrate on the IUE monitoring. In \S~2 we present the 
IUE observations and data analysis, in \S~3 we describe the UV light curves 
and spectral variability, in \S~4 we discuss these results and in \S~5 we
summarize our conclusions.
 
\section{Observations and Data Analysis}

\subsection{Observing Strategy and Spectra Reduction}

IUE was scheduled for continuous observations (three 8-hr shifts per day) from
1994 May 15 to 25 inclusive, with 8 brief ($\simlt$2 hr) and 5 longer (between
4 and 17 hr) gaps due to Earth occultation and to a few time-critical programs.
The target acquisition was done through a double blind offset: first we 
pointed to the nearby bright star SAO 213406 (V = 6.5, at 44$^{\prime}$ 
distance from the source), then to the fainter SAO 213450 (V = 9.2, at 
4.5$^{\prime}$), and finally to the target itself. The SWP and LWP cameras 
were exposed in low dispersion mode alternately for 55 and 25 minutes 
respectively, to achieve 
comparable signal-to-noise ratio
in both cameras, for a typical UV spectral slope of 
PKS~2155--304 ($\alpha_\nu \simeq 1$). In the absence of operational problems,
we obtained one pair of spectra each 96 minutes, due to satellite maneuvering
and camera preparation overheads. This time interval was chosen to phase with 
the ASCA satellite orbital period to allow cleaner cross-correlation analysis 
between the UV and X-ray light curves; depending on the overheads, some of the
spectra had slightly longer or shorter integration times than the nominal
25 and 55 minutes. One long-wavelength spectrum (LWP 28222) and three 
short-wavelength spectra (SWP 50815, 50840, 50854) were very underexposed (the
exposure times were less than half the normal values) and were discarded from 
the subsequent analysis.

The photometrically flat-fielded and geometrically corrected images were
inspected to assure proper target centering. One long-wavelength spectrum was 
unusable because of off-axis placement of the aperture during exposure (LWP 
28187). 

As of late 1992, scattered solar light severely affects the IUE field of view 
(depending on the satellite position with respect to the Sun) and can
significantly compromise the spectrophotometry longwards of $\sim$2700 \ang\ 
(Caplinger 1995, and references therein). The scattered light also precludes 
useful information about the source brightness at optical wavelengths from the
FES, so no FES counts from the source were recorded.

Spectra were extracted from each of the 236 good IUE images using the TOMSIPS
routine (Ayres 1993; Ayres et al. 1995), a modified version  of the 
Signal-Weighted Extraction Technique (SWET; Kinney, Bohlin, \& Neill 1991a). 
The extracted net fluxes were converted to absolute fluxes using calibration 
curves based on SWP and LWP low dispersion spectra of the white dwarf 
WD G191-B2B. No
correction was applied for the sensitivity degradation of the cameras. The 
complete log of the IUE observations is reported in Table 1. Two typical 
spectra from the campaign are shown in Fig. 1 along with the intrinsic error 
distribution of the spectral flux. The best-fit power-law model for the 
continuum is shown as a solid line.

The results of the TOMSIPS extraction were compared with alternative 
processings using the standard IUESIPS, GEX (Urry \& Reichert 1988), and the 
Final Archive NEWSIPS (Nichols \& Linsky 1996). In the first two cases a 
general consistency was found within $\sim$10-15\%, though GEX gave anomalous
results in a few cases. The NEWSIPS LWP spectra from the end of the US2 shift,
which is the most heavily affected by background radiation, turned out to be 
unreliable. For these cases, the signal-to-noise ratio of the spectral data is
so low (because of the high background level) that the extraction technique 
uses only two spline nodes to fit the cross-dispersion profile (Imhoff 1996). 
In presence of
solar light contamination, which ramps up dramatically above 2800 \ang, if 
only two spline nodes are used to fit the cross dispersion profile, the fit 
cannot accurately follow the rapid decline in flux below 2800 \ang, resulting 
in overestimated flux in a large portion of the spectrum.
For the SWP, the NEWSIPS and TOMSIPS light curves at 
1400 \AA\ are in agreement, apart from the different applied calibrations.

The expected interstellar extinction of the UV flux due to Galactic neutral 
hydrogen is $A_V$ = 0.08 mag, corresponding to a column density
N$_{HI} = 1.36 \times 10^{20}$ cm$^{-2}$ (Lockman \& Savage 1995),  
assuming a gas-to-dust ratio 
N$_{HI}/E_{B-V}$ = 5.2$\times 10^{21}$ cm$^{-2}$ mag$^{-1}$ (Shull \& Van 
Steenberg 1985) and a total-to-selective extinction ratio $A_V/E_{B-V} = 3.1$
(Rieke \& Lebofsky 1985). The 236 IUE spectra are well fitted by simple
power-law models plus this assumed reddening (\S~2.2). However, with $A_V$ as 
a free parameter, the $\chi^2$ associated with a power-law model is minimized 
(both over the full set of SWP and over the set of merged SWP+LWP spectral 
flux distributions) for $A_V$ = 0.4 mag, with a high degree of significance 
($>$ 99.99\%, according to the F-test). This value is inconsistent with that 
deduced from the Galactic column density, and in fact with the results from 
U93, who found no such excess reddening. This might indicate variable 
absorption at the source or in intervening material (Bruhweiler et al. 1993). 
Nevertheless, since PKS~2155--304 was observed by EUVE during the campaign 
(which would have been unlikely if the extinction were so high) and since the 
$A_V$ = 0.08 mag fits are still acceptable, we conservatively adopted $A_V$ = 
0.1 mag for consistency with U93. (This was derived, under the same above 
assumptions, from the hydrogen column density determined by the $HI$ 21 cm 
survey of Stark et al. (1992), N$_{HI} = 1.78 \times 10^{20}$ cm$^{-2}$.) Note
that for $A_V$ = 0.1 mag, the SWP+LWP spectral indices are consistent with a 
simple power-law, while $A_V$ = 0.4 mag would require that the SWP slope is 
steeper by $\Delta\alpha \sim 1$ than the LWP (Fig. 2). For the dereddening
corrections we used the extinction curve of Seaton (1979), as in U93. The 
updated dereddening curve of Cardelli, Clayton, \& Mathis (1989) would imply 
an average discrepancy of the fitted parameters not exceeding their 
uncertainties.

\subsection{Spectral Fitting}

For the spectral analysis we followed a procedure similar to U93. Through an 
iterative, chi-squared minimization fitting routine, the dereddened SWP 
spectral flux distributions (1230-1950 \ang) were fitted in wavelength space 
to a simple power-law model of the form $F_\lambda \propto \lambda^{-\beta}$, 
ignoring the regions 1277-1281 \ang, 1286-1290 \ang, 1660-1666 \ang, and 
1780-1800 \ang, which are affected by camera artifacts (Crenshaw, Bruegman, \&
Norman 1990), as well as regions of individual spectra contaminated by cosmic 
ray hits. The fit was normalized to 1560 \ang, which is the flux-weighted
mean wavelength of the chosen interval for a spectral index $\beta = 1$.

Similarly, power-law spectral fits were made to the LWP spectra in the 
2100-2700 \ang\ region, which is not heavily affected by the solar
scattered light. Because the signal in the 2700-2800 \ang\ region did not 
exceed the fit curve extrapolation, and was well represented by the same 
power-law, we concluded that the effect of the IUE baffle anomaly is 
negligible shortward of 2800 \ang\ (as suggested also by direct image 
inspection), so we extended the fitted region to 2800 \ang. The fiducial 
wavelength of the fit normalization, computed as above, was chosen to be 
2580 \ang.

The 1-$\sigma$ uncertainties associated with the fitted fluxes are generally 
less than 1\% for SWP spectra, and a few percent for LWP due to the large 
intrinsic errors affecting the spectral signal between 2100 \AA\ and 2400 \ang.
To these uncertainties a $\sim$1\% photometric error was added in quadrature, 
following U93 and Edelson et al. (1992). The best-fit parameters, energy 
spectral indices $\alpha$ (where $\alpha = 2 - \beta$) and fitted fluxes at
1400 \AA\ and 2800 \ang\ (SWP and LWP respectively), are given in Table 1, as 
are the reduced $\chi^2$ values for each fit ($\chi^2_\nu$). We also fitted a
power-law model to pairs of SWP and LWP spectra taken close together in time. 
The fit was done over the wavelength range 1230-2800 \ang, excluding the same 
regions that were discarded in the SWP spectra and the 1900-2150 \ang\ 
interval, which is affected by large errors. During the first day of 
monitoring, the extremely fast variability does not allow any meaningful match
between SWP and LWP exposures. Therefore, since the determination of the 
spectral index would yield unreliable values, due to lack of simultaneity,
we excluded the first six pairs of spectra from computation of the combined
energy index $\alpha_C$. The combined SWP+LWP fit results are presented in
Table 2. 

The present analysis leads to a steeper average spectral slope than reported 
for the 1991 data ($\langle \alpha_{SWP} \rangle$ is larger by $\sim$28\%, 
 $\langle \alpha_{LWP} \rangle$ by $\sim$14\%, and $\langle \alpha_C \rangle$ 
by $\sim$44\%), which were reduced with the SWET method. We 
investigated the cause of this difference by re-analyzing the 1991 data after 
extracting the spectra with the TOMSIPS routine and found that the change is 
mostly due to the different adopted calibration curves, and to the fact that
the 1991 data were corrected for the SWP camera sensitivity degradation, with
smaller effects due to the more limited fitting range at the long wavelengths
and (only marginally) to the different extraction algorithms (e.g., profile 
normalizations). Fitting the 1991 TOMSIPS extracted spectra to a power-law 
yields average spectral indices that are consistent with those obtained for 
the 1994 data. (See Table 3 for a synoptic comparison between the 1991 and 
1994 sets.)

The average $\chi^2_\nu$ values for the SWP, LWP, and merged spectral fits 
are 1.02, 0.67, and 1.00, respectively. This indicates that, given the derived
flux errors, the power-law model is acceptable, therefore no rescaling was 
applied to the intrinsic flux errors (as had been done in U93). 
To determine errors on spectral index we looked at the differences between 
spectral indices for pairs of adjacent spectra taken closer in time than 0.1 
days, excluding the first six points in each energy index curve (where the 
variability was much faster than the exposure time), leaving 94, 95, and 91 
pairs of SWP, LWP, and merged spectra, respectively.

Assuming that the spectral index does not 
vary significantly between two observations spaced only $\sim$2 hours apart in
time (which is not the case in the first part of the monitoring), the 
difference 
divided by the sum in quadrature of their individual errors should be normally
distributed, with a unity variance. Since we found variances larger than unity
(2.13, 1.17, and 2.86 respectively for the SWP, LWP, and merged spectra), we 
applied, as in U93, a correction to the spectral index errors equivalent to 
such variance. The spectral index distributions for the SWP and LWP overlap 
(though the former is much narrower), indicating that the assumed extinction 
produces a consistent result for both SWP and LWP spectral shapes (see Fig. 
2a).
For $A_V$ = 0.4 mag, the fitted fluxes at 1400 \AA\ and 2800 \AA\ would be 
larger by $\sim$110\% and $\sim$65\%, respectively, and the mean SWP and LWP 
spectral indices would be smaller, the differences being $\sim$0.3 and 
$\sim$1.2, respectively, compared to the values given in Table 1 for $A_V$ = 
0.1 mag. The SWP and LWP spectral index distributions for $A_V$ = 0.4 mag 
(Fig. 2b) differ by $\Delta\alpha \sim 1$, suggesting that such a high 
extinction value is unlikely. 

\section{Results}

\subsection{UV Light Curves}

Both long- and short-wavelength IUE light curves clearly show strong fast 
variability (Fig. 3). The most striking result of the campaign is the high 
amplitude, extremely rapid flux variability detected during the first day of 
monitoring (Fig. 3b). At 2800 \ang, flux variations as large as a factor 2.2 
were found in 1.5 hours (i.e., between one LWP integration and the other), 
implying that significant variability can occur on time scales shorter than 
the typical IUE integration time. 

The flux at 1400 \ang\ also varied remarkably in the first part of the 
monitoring, though with smaller observed amplitude ($\sim$25\% in 1.5 hours). 
This behavior is entirely new and unexpected, since the many previous IUE
observations showed only minor differences in SWP and LWP variability, usually
in the sense of a larger amplitude in SWP. In Fig. 3b the two light curves
are compared normalizing the 2800 \ang\ and the 1400 \ang\ fluxes to their
respective mean values (computed excluding the initial flares, i.e. the first 
six flux points of each curve). The figure shows that the SWP light curve can 
be reconciled with the hypothesis of a complete correlation with the LWP (i.e.,
no spectral variability) given that the SWP integrations are twice as long as
for the LWP and allowing for the fact that the shortest time scale variability
is probably not resolved in either band. 

Because the observed variability at 2800 \ang\ is of unprecedented amplitude 
for such
short time scales, we investigated possible extrinsic causes. First, we asked
whether motion of the spacecraft could have caused a dip in flux as the 
source drifted out and back into the large aperture. We examined carefully
the line-by-line file for the image with the sharpest drop in spectral flux, 
LWP~28142. The spectrum is visible with a good signal-to-noise ratio
in the central 
part of the line-by-line image. This, together with proper centering of the 
adjacent SWP spectra with respect to the geocoronal Ly$_{\alpha}$ line (which 
fills the aperture), exclude a drift of the target along the long dimension 
of the aperture. 
Similarly, a drift along the dispersion direction is excluded because of
the location of the reseau marks at the same wavelengths as in the other 
images, and the 
accuracy of the spacecraft slews between the target and the offset star.
Second, we ruled out a mis-registration of the Optimal extraction slit with
respect to the spectrum, since several independent extraction methods 
(including the original boxcar) give essentially the same spectrum. 
Third, this ``dip'' event might resemble some sudden 
flux drops detected in LWP data during IUE intensive monitorings, which have
been ascribed to voltage failures in the LWP camera 
(LWP anomaly, T. Teays, private communication, 1996). However, 
such spurious effects are very rare, and no firm 
conclusion has been achieved on the correlation between the occurrence of
the phenomenon and electric current depletion during the exposure. 
More importantly, these events are confined to the LWP, while
the event in PKS~2155--304 has a (less dramatic) counterpart in the SWP light
curve, in the EUVE light curve (Marshall et al. 1996), and in polarized 
optical light (Pesce et al. 1996).
Considering all these points, we conclude the rapid initial variations reflect
actual events in the BL Lac object.

In both light curves, after the very active period recorded at the beginning 
of the campaign, the flux stays relatively constant (after May 16), with a 
gentle increase and decline of $\sim$10\% over 3.5 days, and then there is a 
prominent 2-day flare, with a total increase of $\sim$35\%, starting on
May 19. Further variation in the last four days of the monitoring is 
$\simlt$20\% in both bands. Throughout the monitoring, small (5-10\%) and 
rapid ($\sim$ hours) flux variations are superimposed on the more dramatic 
flares.

A variability test (Edelson 1992) yields similar variability indices for the 
SWP and LWP light curves, which are marginally consistent within the 10\% 
error, calculated as in Edelson et al. (1995; $v_i \equiv {\sigma_F\over 
\langle F \rangle}$, where $\sigma_F$ is the standard deviation of the 
flux, and $\langle F \rangle$ its average value):
$v_i^{SWP}$ = 0.11, $v_i^{LWP}$ = 0.13. 
Performing the test on the light curves after removing their first, 
dramatically variable portions, results in a 0.10 variability index both for 
SWP and LWP. This is consistent with the results of the intensive November 
1991 campaign on PKS~2155--304, when comparable amplitude variability was 
found in both IUE wavelength ranges. Systematically larger amplitude
variability is found at shorter UV wavelengths than at longer ones for blazars
observed over longer time scales (Kinney et al. 1991b; Edelson 1992), 
but the difference is not significant given the errors (Treves \& Girardi 
1991), and such differences are not observed on short time scales, where
rapid variations are resolved (see U93).

We computed auto- and cross-correlation functions for the light curves by 
means of the Discrete Correlation Function method (DCF; Edelson \& Krolik 
1988). Since the character of the variability in the first day is extreme and 
unresolved, we systematically removed the first 6 flux points in both light 
curves prior to application of the DCF routine.

The auto-correlation function of the present SWP and LWP light curves (Fig. 4)
does not show any evidence of periodicity (as would be implied by 
characteristic ``humps'' on the auto-correlation function curve at non-zero 
lags) nor is any 
visible in the light curves, in contrast to the previous findings. The 
auto-correlation function of a data train shows features at the timescale
corresponding to recurrence in 
its variations, such as periodicity or quasi-periodicity would imply. This 
allowed U93 to find a quasi-periodic variation in the November 1991 IUE light 
curves of PKS~2155--304 with a $\sim$0.7 days time scale (a strict periodicity
was
shown not to be statistically significant by Edelson et al. 1995). The SWP and
LWP light curves are well correlated with each other, with no apparent lag 
larger than $\sim$0.1 days (Fig. 5), which is the approximate temporal 
resolution.

\subsection{Spectral Shape and Variability}

During the first day of monitoring the UV spectrum of PKS~2155--304 varied
both in the LWP and, less prominently, in the SWP range (Fig. 6). 
After May 16, the slopes of the SWP and LWP spectra have an overall
fractional variability $\sigma_\alpha \over <\alpha>$ of 0.08 and 0.28, 
respectively (with a 10\% uncertainty on these values), according to the 
variability test of Edelson (1992). Testing the spectral index behavior against
a constant trend yields a $\chi^2_\nu$ of 1.8 for SWP and 2.2 for LWP, which 
implies a probability of constancy of less than 0.1\%. The central resolved 
flare is accompanied by spectral variations of more modest amplitude: 
$\chi^2_\nu \sim 2$ is found for both index curves, whereas $\chi^2_\nu$ = 53
and $\chi^2_\nu$ = 21 is associated with the flux variations in the SWP and 
LWP, respectively.

The auto-correlation function of the LWP spectral index shows equally spaced 
peaks with $\sim$1 day separation. This periodicity (also recognizable in Fig.
6b) is spurious, deriving from  the periodic background contamination during 
the US2 shift. The SWP spectral index auto-correlation function does not 
exhibit any significant features. No clear trend is visible between the fluxes
at 1400 \ang\ and the SWP spectral slopes, but their cross-correlation has a 
minimum at a lag of $\sim -1$ day (Fig. 7), which implies that spectral 
flattening (steepening) leads flux increases (decreases). Limiting the 
cross-correlation function computation to the segments of the light and
index curves corresponding to the central flare yields no evidence that the 
effect might be dominated by the behavior during the outburst.
 
The slope $\alpha_C$ of the SWP+LWP spectra is
significantly variable ($v_i$ = 0.08, with $\chi^2_\nu$ = 2.8; see Fig. 6c, 
where $\alpha_C$ is plotted only for observations taken after May 16). During 
the central flare, a $\chi^2_\nu \simeq 3$ is associated with the $\alpha_C$ 
variation. Cross-correlating $\alpha_C$ with the flux at 2000 \ang\ yields the
same results as found for the SWP: spectral hardening $\sim$1 day in advance
of flux rise.

\section{Discussion}

IUE monitoring of PKS~2155--304 in May 1994 has given us the best sampled UV 
light curve ever obtained for an AGN, with 2.2 times the temporal extension of
the intensive part of the November 1991 campaign. On average, the state 
presently detected was $\sim$20\% lower than in 1991 in both the SWP and the 
LWP. The observed flux variability behavior was also remarkably different. The
light curves from the two intensive campaigns are shown for comparison in Fig.
8. In particular, we do not see the quasi-periodicity seen in 1991 (see flux 
auto-correlation functions, Fig. 4), which has therefore to be regarded as 
random or transitory. No significant spectral change is seen between the two 
epochs (see \S~2.2 and Table 3). Compared to the IUE archival data (Edelson et
al. 1992; Pian \& Treves 1993) both UV flux and spectral index were about 
average in May 1994.

The 1994 data exhibit dramatic and unprecedented variability, which is still 
underresolved, during the first day of the monitoring and later a well sampled 
flare of a factor of $\sim$35\% in $\sim$1.5 days, visible in both the 1400 
\ang\ 
and 2800 \ang\ light curves without any significant difference. The flux 
rise to maximum is longer than the fading ($\sim$1 day) toward the previous 
``quiescent'' state (as indicated by the asymmetric shape of the flare, Fig. 
3a), in agreement with the finding that UV flux decrease in blazars is usually
sharper than brightening (Edelson 1992). The longer and ``structured'' rise 
requires that the mechanism producing the flare is not ``instantaneous'' but 
rather intrinsically long or diluted, possibly by light travel time effects in
the emission region, or by multiple smaller events.

The fast fluctuations seen at the beginning of the light curve are truly
exceptional. During the first day the LWP light curve exhibits a variation of 
a factor 2.2 in 8 hours and a second flare of similar amplitude in 1.5 hours 
(Fig. 3b). Rapid variability is simultaneously seen in the SWP and in the LWP.
Fast (though unresolved) variations have also been detected during the optical 
observations simultaneous to the present IUE campaign (Pesce et al. 1996). 
Defining the flux doubling time scale as $\tau = {F_{min} \over 
{F_{max} - F_{min}}} \Delta t$ one obtains $\tau_{SWP}$ = 6.87 hr and 
$\tau_{LWP}$ = 1.36 hr as minimum values at 1400 \AA\ and 2800 \AA\ 
respectively during the monitoring. This is the most rapid observed flux 
doubling for PKS~2155--304 either in the UV or optical range, where typical 
values are of the order of days (see U93; Carini \& Miller 1992; Miller 1996).
Doubling time scales as short as $\sim$1 hour have been seen in PKS~2155--304 
optical light curves (Pesce et al. 1996; Paltani et al. 1996), but for typical
amplitudes much smaller than a factor of 2. In X-rays, 1-hour doubling 
time scales for this source are relatively more common (Morini et al. 1986). 
Similar events were observed in several X-ray bright blazars in the X-ray band
(see e.g., H~0323+022, Feigelson et al. 1986; PKS~2155--304, Treves et al. 
1989, Sembay et al. 1993; PKS~0716+714, Cappi et al. 1994), and also in the
$\gamma$-rays (Mkn~421, Macomb et al. 1995; PKS~1622-297, Mattox et al. 1997). 

Conspicuous spectral variations corresponding to the big central flare are not
seen, a behavior reminiscent of other BL Lacs in optical and UV (OJ~287, 
Pian et al. 1996; Sillanp\"a\"a et al. 1996; PKS~0716+714, Wagner \& Witzel
1994, see however Ghisellini et al. 1996, who find opposite results for this
object in the optical). This result is to be compared with the outcome of 
longer term 
monitorings of blazars (days to years), showing that UV spectral variations
do occur but are generally modest compared to flux changes and often weakly or
not clearly correlated with them (Edelson 1992; Shrader et al. 1994; Koratkar 
et al. 1996). Where correlation is found, it is generally in the sense of a
harder spectrum for brighter flux (Urry et al. 1988; Bonnell et al. 1994),
in qualitative agreement with models based on radiative cooling.

A significant feature of the cross-correlation of flux and spectral index for 
the SWP (Fig. 7) is a minimum at $\sim$ --1 day, which represents 
anti-correlation of the two quantities, consistent with the longer term
spectral variability mentioned above. The negative time-lag means that the 
spectrum is hardening (softening) $\sim$1 day before the flux increases 
(decreases), a result which, as yet unexplained, was also found in the 
November 1991 campaign (see discussion in U93).

The variability in the initial part of the light curve is so rapid that we 
cannot test color variations using the joint LWP and SWP ranges during the 
event. However the spectral index in the LWP range takes one of its lowest
values ($\alpha_\nu$ = 0.5) and its maximum value ($\alpha_\nu$ = 1.7) in 
correspondence to the relative maximum flux observed on May 15.95 and to the 
deep minimum observed on May 16.08, respectively. Thus the little information 
we have points to some spectral variability rather than to an achromatic event.

The completely resolved central flare, together with the $\sim$50\% and 
$\sim$80\% correlated flares detected at the extreme UV and X-ray 
wavelengths, respectively (Marshall et al. 1996; Kii et al. 1996; Urry et al. 
1996), is consistent with a variability amplitude monotonically increasing 
with energy, as expected for a synchrotron flare in an inhomogeneous jet 
(Celotti, Maraschi, \& Treves 1991; Georganopoulos \& Marscher 1996). 

The excellent correlation of the SWP and LWP light curves implies that the UV 
emission in the 1200-3000 \AA\ range is produced within a unique emitting 
region, without difference in the electron cooling times at these wavelengths 
larger than an hour. Since these must be equal to or shorter than the UV 
fading time scales, we can estimate a lower limit for the magnetic field in 
the UV emitting portion of the jet (Blandford 1990), locally  
approximated as a homogeneous region:

$$
B \ge 1.4 \times \nu_{15}^{-1/3} t_{hr}^{-2/3}(\delta/10)^{-1/3} ~~~~~ G, 
$$

\noindent
where $\nu_{15}$ is the frequency in units of $10^{15}$ Hz, $t_{hr}$ is the 
observed variability timescale in hours, and $\delta$ represents the Doppler 
factor of the relativistic bulk motion. Based on the shortest observed 
variability time scale (1.5 hr at 2800 \AA) and assuming $\delta \sim 10$, we 
derive $B \ge 1$ G.

The detection of GeV $\gamma$-rays from PKS~2155--304 (Vestrand, Stacy, \& 
Sreekumar 1996) indicates that inverse Compton radiation due to electrons 
scattering off the synchrotron photons or other soft seed photons is
significant. The seed photons are expected to emit
at optical and UV wavelengths ($\nu_{IC} \sim \gamma^2 \nu_S$, where the 
maximum electron energy $\gamma$ is typically $100-1000$). Given the derived 
lower limit on the magnetic field and the measured 
synchrotron luminosity in the UV ($L_S$), we can estimate the expected 
$\gamma$-ray luminosity ($L_{IC}$) in the homogeneous case of an emitting blob
of radius $R_{blob} = c t_{var} \delta$. We consider the following relation
between the observed synchrotron and inverse Compton luminosities:

$$
{L_{IC} \over L_S} \, \simeq \, {U_S \over U_B} \, =
{L_S \delta^{-4} \over \, {4\pi R_{blob}^2 c}} \cdot {8\pi \over B^2},
$$

hence

$$
L_{IC} = {2 L_S^2 \over {c^3 B^2 t_{var}^2 \delta^6}},
$$

\noindent
where $U_B$ is the magnetic energy density, and $U_S$ is the 
synchrotron radiation energy density. 
For $L_S = 1.2 \times 10^{46}$ erg s$^{-1}$, corresponding to the UV
emission alone, and $B \ge 1$ G,
$t_{var} = 1.5$ hr, $\delta \sim 10$, one obtains 
$L_{IC} \le 3.4 \times 10^{47}$ erg s$^{-1}$, consistent with the observed 
$\gamma$-ray luminosity of $2.5 \times 10^{45}$ erg s$^{-1}$. 
The only quantities not directly observed are $B$ and $\delta$. The
approximation $\delta \sim 10$ is supported by other observations; given the 
$\gamma$-ray limit here, $B$ could be an order of magnitude larger than
1 G.

The extremely rapid variation observed at the beginning of the UV monitoring
implies that the limit $\Delta L/\Delta t = 
2 \times 10^{42} \eta$ erg s$^{-2}$ (Fabian 1979) is slightly exceeded (by a
factor of $\sim$1.2) if an accretion efficiency $\eta = 0.1$ is assumed,
for a redshift $z$ = 0.116 (Falomo et al. 1993), a Hubble constant
$H_0 = 50$ km s$^{-1}$ Mpc$^{-1}$ and a deceleration parameter $q_0$ = 0.5, 
and hence the rapid variations support the idea of relativistic beaming. Based
only on the flux change detected in the 2100-2800 \AA\ band, the beaming 
factor need not exceed unity, but it might be well larger considering the 
simultaneous (and in some cases larger) flux variation in other parts of the 
electromagnetic spectrum, which is unfortunately undersampled (see Urry et al.
1996). The observed time scale corresponds to a very small emission region, 
only $\sim 10^{14}$ cm if beaming corrections are not applied, or 
$\sim 10^{15}$ cm for 
$\delta\simeq 10$. Notice that the constraint on the presence of beaming 
derived only from the variability in the UV band is a factor of 4 more 
stringent than found in U93, but weaker than that determined by Morini et al. 
(1986), who detected a rapid increase in X-ray (1-6 keV) flux for which 
$\Delta L/\Delta t$ exceeded the above limit by a factor 10. 

\section{Conclusion}

The May 1994 IUE monitoring on PKS~2155--304, which was part of a simultaneous
multiwavelength campaign from radio to X-rays, yielded the best sampled UV 
light curve for this or any blazar and revealed significant flux variability 
at different time scales, from hours to days. Spectral changes are generally 
modest and not clearly correlated with flux variations. A resolved central 
flare of $\sim$35\% amplitude was observed in both IUE cameras, and was likely 
correlated with flares of different amplitude and duration at higher energies 
(Urry et al. 1996). The 1.5-hour flux variation of a factor 2.2 seen at 2800 
\AA\ during the first day of monitoring is unprecedented for blazars as a 
class. This event, which suggests the occurrence of variability on time
scales even shorter than the IUE time resolution, represents a definite 
violation of the limits on luminosity variability, therefore implying the 
presence of relativistic beaming. Both variability events are interpreted 
within a scenario in which synchrotron radiation is the primary emission 
mechanism, and a lower limit of 1 G on the intensity of the magnetic field is 
determined, which turns out to be consistent with a calculation of this 
physical quantity based on the multiwavelength data and with the $\gamma$-ray 
flux observed by EGRET.

Comparison with the results from the 1991 IUE intensive monitoring of
PKS~2155--304 shows that the variability characteristics detected at the two 
epochs are fundamentally different. In particular, unlike the findings of the
1991 campaign, no periodicity is seen in the present data. The $\sim$1 day 
anti-correlation between flux and spectral variations is however maintained. 

The fastest variability in the UV has been 
largely undersampled with IUE for PKS~2155--304, and in general for other 
blazars. UV telescopes allowing a better time resolution and continuous
coverage are clearly needed. 
Apart from HST, which has rarely been dedicated to long monitorings, none are
presently available. While the fast sporadic variability may be most apparent 
in the UV, it should be expected to manifest itself, possibly in a less extreme
form, also at optical wavelengths. Some indication is already present in the 
optical data of the 1994 campaign, especially in the polarization measurements
(Pesce et al. 1996). These arguments point to the importance of intensive, 
systematic monitoring in the optical band, possibly with polarization 
information. Such programs, which have led to the discovery of intraday 
variability in a number of blazars (Wagner \& Witzel 1995; Miller 1996; 
Sillanp\"a\"a et al. 1996; Smith 1996), can be carried out with medium sized 
telescopes and standard instrumentation, but need long, uninterrupted 
observing runs, possibly coordinated among different sites.

\acknowledgements 

EP, CMU, JEP, and PG acknowledge support from NASA grants NAG5-1034 and
NAG5-2499; EP acknowledges support from a NATO-CNR Advanced Fellowship;
RIK acknowledges support from NASA LTSA grant NAGW-2120.
We are grateful to T. Ayres for clarification of the TOMSIPS routine and
to P. Smith for critical comments. The IUE staff at Vilspa and GSFC is 
acknowledged, and particularly we would like to thank R. Arquilla, R. Bradley,
J. Caplinger, D. De Martino, M. England, C. Gonzalez, A. Groebner,  C. Imhoff,
N. Loiseau, C. Loomis, D. Luthermoser, B. McCollum, J. Nichols, R. Pitts, 
C. Proffitt, L. Rawley, P. Rodriguez, M. Schlegel, L. Taylor, T. Teays,
and R. Thompson for assistance with IUE observations and data reduction.
C. Bailyn, 
R. Bohlin,
J. Bregman,
W. Brinkmann, 
M. Carini, 
L. Chiappetti,
M. Donahue, 
E. Feigelson,
A. Fruscione, 
A. K\"onigl, 
L. Kedzior, 
Y. Kondo,
A. Koratkar,
J. Krolik,
A. Lawrence,
F. Makino,
P. Martin, 
H. Miller,
P. O'Brien, 
G. Reichert,
A. Sadun,
M. Sitko, 
P. Smith,
A. Szymkowiak,
G. Tagliaferri, 
E. Tanzi,
S. Wagner, 
R. Warwick,
A. Wehrle are acknowledged for their support to the observing project.


%
%
\begin{center}
\begin{tabular}{ccccccccc}
\multicolumn{9}{c}{{\bf Table 1:} Log of IUE Observations of PKS~2155--304 and 
}\\
\multicolumn{9}{c}{Power-Law Fit Parameters of Dereddened Spectra}\\
\hline
\hline
IUE Image & Observation & Exposure  & Observatory & $F_\nu^a$ & 
$\sigma_{F_\nu}$ & $\alpha_\nu^b$ & $\sigma_{\alpha_\nu}$ & $\chi^2_\nu$ \\
& Midpoint (UT) & Time & (Goddard & (mJy) & (mJy) & & & \\
& (day of May 94) & (min) & or Vilspa) & & & & &\\
\hline 
 SWP 50773 & 15.71418 & 60 & G &   7.03 &  0.13 &  0.95 &  0.10 &  0.66 \\  
 SWP 50774 & 15.78942 & 55 & G &   7.75 &  0.12 &  0.88 &  0.07 &  0.81 \\  
 SWP 50775 & 15.85629 & 35 & G &   8.66 &  0.13 &  1.19 &  0.06 &  0.88 \\  
 SWP 50776 & 15.98259 & 45 & V &   8.95 &  0.13 &  1.04 &  0.05 &  0.79 \\  
 SWP 50777 & 16.05153 & 44 & V &   8.63 &  0.13 &  1.04 &  0.06 &  1.02 \\  
 SWP 50778 & 16.11782 & 55 & V &   8.01 &  0.12 &  0.88 &  0.05 &  0.89 \\  
 SWP 50779 & 16.18512 & 55 & V &   9.89 &  0.14 &  0.90 &  0.05 &  1.02 \\  
 SWP 50780 & 16.25361 & 55 & V &   9.92 &  0.14 &  0.98 &  0.05 &  0.97 \\  
 SWP 50781 & 16.32059 & 50 & G &   9.90 &  0.14 &  0.98 &  0.05 &  0.96 \\  
 SWP 50782 & 16.38535 & 55 & G &   9.89 &  0.14 &  0.94 &  0.05 &  0.88 \\  
 SWP 50783 & 16.45285 & 55 & G &   9.91 &  0.14 &  1.00 &  0.05 &  1.06 \\  
 SWP 50784 & 16.51876 & 55 & G &   9.86 &  0.14 &  1.01 &  0.05 &  1.05 \\  
 SWP 50785 & 16.58534 & 55 & G &   9.91 &  0.14 &  0.94 &  0.05 &  1.19 \\  
 SWP 50786 & 16.65740 & 42 & G &  10.19 &  0.16 &  0.92 &  0.08 &  0.87 \\  
 SWP 50787 & 16.71852 & 55 & G &   9.72 &  0.34 &  0.86 &  0.25 &  0.36 \\  
 SWP 50788 & 16.78274 & 45 & G &   9.85 &  0.17 &  1.11 &  0.09 &  0.65 \\  
 SWP 50789 & 16.84537 & 40 & G &  10.34 &  0.15 &  0.94 &  0.05 &  0.82 \\  
 SWP 50790 & 16.98689 & 48 & V &  10.11 &  0.14 &  1.00 &  0.05 &  1.08 \\  
 SWP 50791 & 17.05113 & 55 & V &  10.15 &  0.14 &  0.94 &  0.05 &  1.39 \\  
 SWP 50792 & 17.11719 & 55 & V &  10.17 &  0.14 &  0.98 &  0.05 &  1.06 \\  
\hline  
\end{tabular}
\end{center}
\begin{center}
\begin{tabular}{ccccccccc}
\multicolumn{9}{c}{{\bf Table 1:} {\it - continued.}} \\
\hline
\hline
IUE Image & Observation & Exposure  & Observatory & $F_\nu^a$ & 
$\sigma_{F_\nu}$ & $\alpha_\nu^b$ & $\sigma_{\alpha_\nu}$ & $\chi^2_\nu$ \\
& Midpoint (UT) & Time & (Goddard & (mJy) & (mJy) & & & \\
& (day of May 94) & (min) & or Vilspa) & & & & &\\
\hline 
 SWP 50793 & 17.18391 & 55 & V &  10.05 &  0.14 &  1.05 &  0.05 &  1.02 \\  
 SWP 50794 & 17.25061 & 55 & V &  10.21 &  0.14 &  0.97 &  0.05 &  1.09 \\  
 SWP 50795 & 17.31883 & 55 & G &  10.37 &  0.15 &  0.95 &  0.05 &  1.08 \\  
 SWP 50796 & 17.38483 & 55 & G &  10.24 &  0.14 &  1.03 &  0.05 &  1.07 \\  
 SWP 50797 & 17.45105 & 53 & G &  10.33 &  0.14 &  1.00 &  0.05 &  1.26 \\  
 SWP 50798 & 17.51777 & 55 & G &  10.25 &  0.14 &  1.00 &  0.05 &  1.27 \\  
 SWP 50799 & 17.58441 & 55 & G &  10.42 &  0.15 &  0.98 &  0.05 &  1.09 \\  
 SWP 50800 & 17.65082 & 55 & G &   9.96 &  0.15 &  1.01 &  0.07 &  0.81 \\  
 SWP 50801 & 17.71312 & 40 & G &  10.02 &  0.23 &  0.89 &  0.15 &  0.68 \\  
 SWP 50802 & 17.77571 & 30 & G &  10.03 &  0.20 &  1.19 &  0.12 &  0.70 \\  
 SWP 50803 & 17.84478 & 35 & G &   9.89 &  0.14 &  1.07 &  0.06 &  1.09 \\  
 SWP 50804 & 17.97949 & 40 & V &   9.59 &  0.14 &  1.08 &  0.06 &  0.91 \\  
 SWP 50805 & 18.04629 & 55 & V &   9.93 &  0.14 &  1.00 &  0.05 &  1.31 \\  
 SWP 50806 & 18.10763 & 40 & V &   9.61 &  0.14 &  1.06 &  0.06 &  1.04 \\  
 SWP 50808 & 18.38371 & 55 & G &   9.56 &  0.14 &  1.09 &  0.05 &  1.00 \\  
 SWP 50809 & 18.44950 & 57 & G &   9.50 &  0.13 &  1.16 &  0.05 &  1.08 \\  
 SWP 50810 & 18.51637 & 55 & G &   9.44 &  0.13 &  1.14 &  0.05 &  1.05 \\  
 SWP 50811 & 18.58464 & 53 & G &   9.46 &  0.14 &  1.02 &  0.05 &  1.10 \\  
 SWP 50812 & 18.65004 & 55 & G &   9.40 &  0.15 &  0.92 &  0.07 &  0.84 \\  
 SWP 50813 & 18.70978 & 35 & G &   9.08 &  0.21 &  1.03 &  0.15 &  0.70 \\  
\hline  
\end{tabular}
\end{center}
\begin{center}
\begin{tabular}{ccccccccc}
\multicolumn{9}{c}{{\bf Table 1:} {\it - continued.}} \\
\hline
\hline
IUE Image & Observation & Exposure  & Observatory & $F_\nu^a$ & 
$\sigma_{F_\nu}$ & $\alpha_\nu^b$ & $\sigma_{\alpha_\nu}$ & $\chi^2_\nu$ \\
& Midpoint (UT) & Time & (Goddard & (mJy) & (mJy) & & & \\
& (day of May 94) & (min) & or Vilspa) & & & & &\\
\hline 
 SWP 50814 & 18.77382 & 33 & G &   9.59 &  0.19 &  0.85 &  0.12 &  0.71 \\  
 SWP 50815$^c$&19.00072&17 & V &   7.75 &  0.14 &  1.36 &  0.10 &  0.98 \\  
 SWP 50816 & 19.05049 & 55 & V &   9.28 &  0.13 &  1.09 &  0.05 &  0.93 \\  
 SWP 50817 & 19.11502 & 50 & V &   9.23 &  0.13 &  1.06 &  0.05 &  1.03 \\  
 SWP 50818 & 19.18376 & 55 & V &   9.39 &  0.13 &  1.03 &  0.05 &  1.11 \\  
 SWP 50819 & 19.24659 & 47 & V &   9.51 &  0.14 &  1.00 &  0.06 &  1.14 \\  
 SWP 50820 & 19.32086 & 42 & G &   9.22 &  0.14 &  1.03 &  0.06 &  0.93 \\  
 SWP 50821 & 19.38240 & 58 & G &   9.49 &  0.13 &  0.96 &  0.05 &  1.02 \\  
 SWP 50822 & 19.44929 & 57 & G &   9.38 &  0.13 &  1.03 &  0.05 &  0.85 \\  
 SWP 50823 & 19.51545 & 56 & G &   9.68 &  0.14 &  0.97 &  0.05 &  1.07 \\  
 SWP 50824 & 19.58221 & 58 & G &   9.76 &  0.14 &  0.95 &  0.05 &  1.30 \\  
 SWP 50825 & 19.64909 & 54 & G &   9.59 &  0.15 &  0.81 &  0.07 &  0.66 \\  
 SWP 50826 & 19.70796 & 33 & G &   9.20 &  0.23 &  0.94 &  0.17 &  0.56 \\  
 SWP 50827 & 19.77311 & 40 & G &   9.37 &  0.18 &  1.12 &  0.11 &  0.84 \\  
 SWP 50828 & 19.84450 & 40 & G &   9.51 &  0.14 &  0.96 &  0.05 &  1.06 \\  
 SWP 50829 & 19.98050 & 43 & V &  10.04 &  0.14 &  1.01 &  0.05 &  1.15 \\  
 SWP 50830 & 20.04891 & 48 & V &  10.45 &  0.15 &  0.91 &  0.05 &  0.85 \\  
 SWP 50831 & 20.11530 & 53 & V &  10.72 &  0.15 &  0.98 &  0.05 &  0.97 \\  
 SWP 50832 & 20.18150 & 51 & V &  10.95 &  0.15 &  0.99 &  0.05 &  0.94 \\  
 SWP 50833 & 20.24868 & 53 & V &  11.12 &  0.16 &  0.91 &  0.05 &  1.06 \\  
\hline  
\end{tabular}
\end{center}
\begin{center}
\begin{tabular}{ccccccccc}
\multicolumn{9}{c}{{\bf Table 1:} {\it - continued.}} \\
\hline
\hline
IUE Image & Observation & Exposure  & Observatory & $F_\nu^a$ & 
$\sigma_{F_\nu}$ & $\alpha_\nu^b$ & $\sigma_{\alpha_\nu}$ & $\chi^2_\nu$ \\
& Midpoint (UT) & Time & (Goddard & (mJy) & (mJy) & & & \\
& (day of May 94) & (min) & or Vilspa) & & & & &\\
\hline 
 SWP 50834 & 20.31591 & 55 & G &  11.05 &  0.16 &  0.88 &  0.05 &  1.24 \\  
 SWP 50835 & 20.38177 & 57 & G &  11.16 &  0.16 &  0.92 &  0.05 &  1.34 \\  
 SWP 50836 & 20.44870 & 58 & G &  11.03 &  0.15 &  1.09 &  0.05 &  1.31 \\  
 SWP 50837 & 20.51553 & 57 & G &  11.23 &  0.16 &  0.95 &  0.05 &  1.08 \\  
 SWP 50838 & 20.58183 & 56 & G &  11.37 &  0.16 &  1.01 &  0.05 &  1.39 \\  
 SWP 50839 & 20.64849 & 54 & G &  10.69 &  0.17 &  1.06 &  0.08 &  0.66 \\  
 SWP 50840$^c$&20.70360&23 & G &  10.99 &  0.29 &  1.22 &  0.18 &  0.61 \\  
 SWP 50841 & 20.77465 & 35 & G &  11.62 &  0.23 &  0.93 &  0.12 &  0.63 \\  
 SWP 50842 & 20.84296 & 40 & G &  12.47 &  0.17 &  1.03 &  0.05 &  1.16 \\  
 SWP 50843 & 20.98156 & 50 & V &  12.44 &  0.17 &  1.05 &  0.04 &  1.29 \\  
 SWP 50844 & 21.04835 & 55 & V &  12.68 &  0.17 &  1.05 &  0.04 &  1.00 \\  
 SWP 50845 & 21.11513 & 55 & V &  12.66 &  0.17 &  1.00 &  0.04 &  1.36 \\  
 SWP 50846 & 21.18191 & 55 & V &  12.87 &  0.18 &  0.96 &  0.04 &  0.88 \\  
 SWP 50847 & 21.24768 & 52 & V &  12.90 &  0.18 &  0.94 &  0.05 &  1.19 \\  
 SWP 50848 & 21.31563 & 55 & G &  12.82 &  0.18 &  0.98 &  0.04 &  1.14 \\  
 SWP 50849 & 21.38151 & 53 & G &  12.48 &  0.17 &  1.04 &  0.05 &  1.15 \\  
 SWP 50850 & 21.44734 & 55 & G &  12.39 &  0.17 &  0.99 &  0.05 &  1.40 \\  
 SWP 50851 & 21.51433 & 56 & G &  12.60 &  0.17 &  0.92 &  0.04 &  1.30 \\  
 SWP 50852 & 21.58095 & 54 & G &  12.49 &  0.18 &  0.90 &  0.05 &  0.91 \\  
 SWP 50853 & 21.64798 & 52 & G &  10.98 &  0.18 &  1.15 &  0.08 &  0.82 \\  
\hline 
\end{tabular}
\end{center}
\begin{center}
\begin{tabular}{ccccccccc}
\multicolumn{9}{c}{{\bf Table 1:} {\it - continued.}} \\
\hline
\hline
IUE Image & Observation & Exposure  & Observatory & $F_\nu^a$ & 
$\sigma_{F_\nu}$ & $\alpha_\nu^b$ & $\sigma_{\alpha_\nu}$ & $\chi^2_\nu$ \\
& Midpoint (UT) & Time & (Goddard & (mJy) & (mJy) & & & \\
& (day of May 94) & (min) & or Vilspa) & & & & &\\
\hline 
 SWP 50854$^c$&21.70336&23 & G &  10.81 &  0.29 &  1.03 &  0.18 &  0.71 \\  
 SWP 50855 & 21.77585 & 35 & G &  10.51 &  0.21 &  1.13 &  0.12 &  0.65 \\  
 SWP 50856 & 21.84078 & 30 & G &   9.99 &  0.15 &  1.16 &  0.06 &  1.38 \\  
 SWP 50857 & 21.98225 & 50 & V &   9.72 &  0.14 &  1.08 &  0.05 &  1.18 \\  
 SWP 50858 & 22.04740 & 53 & V &   9.61 &  0.14 &  1.06 &  0.05 &  1.09 \\  
 SWP 50859 & 22.11338 & 45 & V &   9.01 &  0.13 &  1.14 &  0.06 &  0.95 \\  
 SWP 50861 & 22.38183 & 55 & G &   9.05 &  0.13 &  1.19 &  0.05 &  0.81 \\  
 SWP 50862 & 22.44815 & 55 & G &   8.95 &  0.13 &  1.09 &  0.05 &  1.04 \\  
 SWP 50863 & 22.51313 & 55 & G &   9.13 &  0.13 &  1.06 &  0.05 &  1.40 \\  
 SWP 50864 & 22.57973 & 55 & G &   9.33 &  0.13 &  0.98 &  0.05 &  0.90 \\  
 SWP 50865 & 22.64640 & 55 & G &   8.83 &  0.15 &  0.98 &  0.09 &  0.68 \\  
 SWP 50866 & 22.70536 & 35 & G &   8.87 &  0.20 &  1.08 &  0.15 &  0.58 \\  
 SWP 50867 & 22.77708 & 48 & G &   9.00 &  0.16 &  0.97 &  0.09 &  0.64 \\  
 SWP 50868 & 22.83884 & 30 & G &   8.61 &  0.13 &  1.14 &  0.07 &  1.14 \\  
 SWP 50869 & 22.98010 & 52 & V &   9.71 &  0.14 &  1.00 &  0.05 &  1.32 \\  
 SWP 50870 & 23.04433 & 45 & V &   9.78 &  0.14 &  1.08 &  0.05 &  1.17 \\  
 SWP 50871 & 23.11241 & 55 & V &   9.85 &  0.14 &  1.03 &  0.05 &  1.66 \\  
 SWP 50872 & 23.18152 & 53 & V &   9.83 &  0.14 &  1.06 &  0.05 &  1.05 \\  
 SWP 50873 & 23.24588 & 50 & V &   9.73 &  0.14 &  1.06 &  0.05 &  1.10 \\  
 SWP 50874 & 23.31339 & 55 & G &   8.90 &  0.13 &  1.19 &  0.05 &  1.34 \\  
\hline  
\end{tabular}
\end{center}
\begin{center}
\begin{tabular}{ccccccccc}
\multicolumn{9}{c}{{\bf Table 1:} {\it - continued.}} \\
\hline
\hline
IUE Image & Observation & Exposure  & Observatory & $F_\nu^a$ & 
$\sigma_{F_\nu}$ & $\alpha_\nu^b$ & $\sigma_{\alpha_\nu}$ & $\chi^2_\nu$ \\
& Midpoint (UT) & Time & (Goddard & (mJy) & (mJy) & & & \\
& (day of May 94) & (min) & or Vilspa) & & & & &\\
\hline 
 SWP 50875 & 23.37897 & 55 & G &   8.88 &  0.13 &  1.16 &  0.05 &  1.01 \\  
 SWP 50876 & 23.44587 & 55 & G &   9.31 &  0.13 &  1.12 &  0.05 &  1.12 \\  
 SWP 50877 & 23.51219 & 55 & G &   9.71 &  0.14 &  1.01 &  0.05 &  1.03 \\  
 SWP 50878 & 23.57843 & 55 & G &   9.85 &  0.14 &  1.02 &  0.05 &  1.02 \\  
 SWP 50879 & 23.64552 & 55 & G &   9.04 &  0.15 &  1.03 &  0.08 &  0.88 \\  
 SWP 50880 & 23.70677 & 40 & G &   8.53 &  0.20 &  1.23 &  0.15 &  0.60 \\  
 SWP 50881 & 23.77962 & 52 & G &   9.06 &  0.15 &  0.97 &  0.08 &  1.07 \\  
 SWP 50882 & 23.83647 & 30 & G &   8.58 &  0.13 &  1.27 &  0.06 &  1.16 \\  
 SWP 50883 & 23.97883 & 55 & V &   9.99 &  0.14 &  1.05 &  0.05 &  1.24 \\  
 SWP 50884 & 24.04752 & 55 & V &   8.96 &  0.13 &  1.05 &  0.05 &  1.13 \\  
 SWP 50885 & 24.11168 & 55 & V &  10.07 &  0.14 &  1.04 &  0.05 &  1.16 \\  
 SWP 50886 & 24.17872 & 55 & V &  10.30 &  0.14 &  1.00 &  0.05 &  1.17 \\  
 SWP 50887 & 24.24718 & 50 & V &  10.18 &  0.14 &  1.03 &  0.05 &  1.14 \\  
 SWP 50889 & 24.71008 & 55 & G &  11.31 &  0.16 &  1.02 &  0.05 &  1.33 \\  
 SWP 50890 & 24.77687 & 55 & G &  10.87 &  0.15 &  1.07 &  0.04 &  1.16 \\  
 SWP 50891 & 24.83460 & 30 & G &  10.51 &  0.15 &  1.09 &  0.06 &  1.11 \\  
 SWP 50894 & 25.71048 & 55 & G &  11.13 &  0.16 &  0.98 &  0.05 &  1.13 \\  
 SWP 50895 & 25.77088 & 37 & G &  11.09 &  0.16 &  1.07 &  0.06 &  0.92 \\  
\hline  
\end{tabular}
\end{center}
\begin{center}
\begin{tabular}{ccccccccc}
\multicolumn{9}{c}{{\bf Table 1:} {\it - continued.}} \\
\hline
\hline
IUE Image & Observation & Exposure  & Observatory & $F_\nu^a$ & 
$\sigma_{F_\nu}$ & $\alpha_\nu^b$ & $\sigma_{\alpha_\nu}$ & $\chi^2_\nu$ \\
& Midpoint (UT) & Time & (Goddard & (mJy) & (mJy) & & & \\
& (day of May 94) & (min) & or Vilspa) & & & & &\\
\hline 
 LWP 28137 & 15.67458 & 30 & G &   7.75 &  0.27 &  0.39 &  0.35 &  0.66 \\  
 LWP 28138 & 15.75657 & 30 & G &  12.04 &  0.37 &  0.81 &  0.29 &  0.57 \\  
 LWP 28139 & 15.83287 & 25 & G &  12.07 &  0.28 &  0.97 &  0.21 &  0.55 \\  
 LWP 28140 & 15.95287 & 25 & V &  17.68 &  0.34 &  0.53 &  0.15 &  0.66 \\  
 LWP 28141 & 16.02132 & 25 & V &  15.12 &  0.32 &  1.55 &  0.18 &  0.63 \\  
 LWP 28142 & 16.08396 & 25 & V &   7.87 &  0.24 &  1.72 &  0.30 &  0.65 \\  
 LWP 28143 & 16.15366 & 25 & V &  17.54 &  0.34 &  1.04 &  0.15 &  0.52 \\  
 LWP 28144 & 16.22040 & 25 & V &  17.30 &  0.33 &  0.84 &  0.15 &  0.60 \\  
 LWP 28145 & 16.28846 & 25 & G &  18.23 &  0.36 &  1.10 &  0.16 &  0.63 \\  
 LWP 28146 & 16.35406 & 25 & G &  17.72 &  0.34 &  0.82 &  0.15 &  0.71 \\  
 LWP 28147 & 16.42109 & 25 & G &  18.02 &  0.36 &  1.24 &  0.16 &  0.67 \\  
 LWP 28148 & 16.48764 & 25 & G &  17.43 &  0.34 &  1.01 &  0.16 &  0.59 \\  
 LWP 28149 & 16.55419 & 25 & G &  17.96 &  0.36 &  0.93 &  0.16 &  0.67 \\  
 LWP 28150 & 16.62071 & 25 & G &  18.54 &  0.37 &  0.70 &  0.16 &  0.74 \\  
 LWP 28151 & 16.68728 & 25 & G &  18.94 &  0.49 &  0.90 &  0.23 &  0.64 \\  
 LWP 28152 & 16.75374 & 25 & G &  18.37 &  0.60 &  0.85 &  0.32 &  0.77 \\  
 LWP 28153 & 16.81868 & 20 & G &  18.94 &  0.40 &  1.23 &  0.18 &  0.63 \\  
 LWP 28154 & 16.95624 & 25 & V &  17.91 &  0.35 &  1.21 &  0.15 &  0.77 \\  
 LWP 28155 & 17.01958 & 25 & V &  18.89 &  0.36 &  1.19 &  0.15 &  0.59 \\  
 LWP 28156 & 17.08590 & 25 & V &  18.33 &  0.35 &  1.18 &  0.15 &  0.87 \\  
\hline  
\end{tabular}
\end{center}
\begin{center}
\begin{tabular}{ccccccccc}
\multicolumn{9}{c}{{\bf Table 1:} {\it - continued.}} \\
\hline
\hline
IUE Image & Observation & Exposure  & Observatory & $F_\nu^a$ & 
$\sigma_{F_\nu}$ & $\alpha_\nu^b$ & $\sigma_{\alpha_\nu}$ & $\chi^2_\nu$ \\
& Midpoint (UT) & Time & (Goddard & (mJy) & (mJy) & & & \\
& (day of May 94) & (min) & or Vilspa) & & & & &\\
\hline 
 LWP 28157 & 17.15254 & 25 & V &  18.09 &  0.34 &  1.09 &  0.15 &  0.75 \\  
 LWP 28158 & 17.21922 & 25 & V &  18.37 &  0.35 &  1.15 &  0.15 &  0.70 \\  
 LWP 28159 & 17.28583 & 25 & G &  18.61 &  0.35 &  0.98 &  0.15 &  0.63 \\  
 LWP 28160 & 17.35373 & 25 & G &  18.27 &  0.35 &  0.98 &  0.15 &  0.62 \\  
 LWP 28161 & 17.42004 & 25 & G &  18.67 &  0.36 &  1.21 &  0.15 &  0.73 \\  
 LWP 28162 & 17.48654 & 25 & G &  18.58 &  0.35 &  0.95 &  0.15 &  0.66 \\  
 LWP 28163 & 17.55325 & 25 & G &  18.67 &  0.36 &  0.91 &  0.15 &  0.61 \\  
 LWP 28164 & 17.61939 & 25 & G &  19.01 &  0.38 &  0.76 &  0.16 &  0.79 \\  
 LWP 28165 & 17.68635 & 25 & G &  18.74 &  0.54 &  0.91 &  0.27 &  0.76 \\  
 LWP 28166 & 17.75058 & 20 & G &  18.85 &  0.70 &  0.58 &  0.36 &  0.65 \\  
 LWP 28167 & 17.81951 & 25 & G &  18.99 &  0.38 &  0.84 &  0.16 &  0.66 \\  
 LWP 28168 & 17.95336 & 25 & V &  18.43 &  0.35 &  1.01 &  0.15 &  0.56 \\  
 LWP 28169 & 18.01455 & 25 & V &  18.50 &  0.35 &  1.24 &  0.15 &  0.70 \\  
 LWP 28170 & 18.08130 & 25 & V &  18.17 &  0.35 &  1.25 &  0.15 &  0.68 \\  
 LWP 28171 & 18.35250 & 25 & G &  17.31 &  0.34 &  0.91 &  0.16 &  0.62 \\  
 LWP 28172 & 18.41790 & 25 & G &  17.72 &  0.35 &  1.00 &  0.16 &  0.66 \\  
 LWP 28173 & 18.48543 & 25 & G &  17.28 &  0.35 &  1.09 &  0.16 &  0.67 \\  
 LWP 28174 & 18.55439 & 25 & G &  17.40 &  0.34 &  0.97 &  0.16 &  0.65 \\  
 LWP 28175 & 18.61839 & 25 & G &  17.89 &  0.37 &  1.00 &  0.17 &  0.75 \\  
 LWP 28176 & 18.68376 & 20 & G &  17.81 &  0.54 &  0.54 &  0.29 &  0.64 \\  
\hline  
\end{tabular}
\end{center}
\begin{center}
\begin{tabular}{ccccccccc}
\multicolumn{9}{c}{{\bf Table 1:} {\it - continued.}} \\
\hline
\hline
IUE Image & Observation & Exposure  & Observatory & $F_\nu^a$ & 
$\sigma_{F_\nu}$ & $\alpha_\nu^b$ & $\sigma_{\alpha_\nu}$ & $\chi^2_\nu$ \\
& Midpoint (UT) & Time & (Goddard & (mJy) & (mJy) & & & \\
& (day of May 94) & (min) & or Vilspa) & & & & &\\
\hline 
 LWP 28177 & 18.74965 & 20 & G &  16.62 &  0.60 & --0.13 &  0.35 &  0.55 \\  
 LWP 28183 & 19.01873 & 25 & V &  17.09 &  0.34 &  1.24 &  0.16 &  0.63 \\  
 LWP 28184 & 19.08539 & 25 & V &  17.10 &  0.34 &  1.22 &  0.16 &  0.60 \\  
 LWP 28185 & 19.15266 & 25 & V &  16.75 &  0.33 &  1.10 &  0.16 &  0.66 \\  
 LWP 28186 & 19.21819 & 25 & V &  16.81 &  0.33 &  0.83 &  0.16 &  0.59 \\  
 LWP 28188 & 19.35045 & 25 & G &  16.78 &  0.33 &  0.79 &  0.16 &  0.59 \\  
 LWP 28189 & 19.41771 & 25 & G &  16.89 &  0.34 &  0.90 &  0.16 &  0.58 \\  
 LWP 28190 & 19.48420 & 25 & G &  16.67 &  0.33 &  1.03 &  0.16 &  0.60 \\  
 LWP 28191 & 19.54995 & 25 & G &  17.36 &  0.34 &  1.08 &  0.16 &  0.57 \\  
 LWP 28192 & 19.61765 & 25 & G &  16.86 &  0.34 &  0.41 &  0.16 &  0.63 \\  
 LWP 28193 & 19.68391 & 25 & G &  16.82 &  0.52 &  0.34 &  0.29 &  0.59 \\  
 LWP 28194 & 19.74867 & 23 & G &  17.99 &  0.75 &  1.17 &  0.42 &  0.65 \\  
 LWP 28195 & 19.81773 & 18 & G &  18.05 &  0.36 &  0.79 &  0.16 &  0.66 \\  
 LWP 28196 & 19.95100 & 25 & G &  17.67 &  0.34 &  0.88 &  0.15 &  0.60 \\  
 LWP 28197 & 20.01746 & 25 & V &  18.56 &  0.35 &  1.02 &  0.15 &  0.71 \\  
 LWP 28198 & 20.08449 & 25 & V &  18.88 &  0.36 &  1.00 &  0.15 &  0.64 \\  
 LWP 28199 & 20.15042 & 25 & V &  18.81 &  0.35 &  1.04 &  0.15 &  0.63 \\  
 LWP 28200 & 20.21753 & 25 & V &  19.57 &  0.37 &  0.91 &  0.15 &  0.61 \\  
 LWP 28201 & 20.28340 & 25 & V &  19.26 &  0.37 &  0.71 &  0.15 &  0.64 \\  
 LWP 28202 & 20.35010 & 25 & G &  19.69 &  0.38 &  0.82 &  0.15 &  0.66 \\  
\hline  
\end{tabular}
\end{center}
\begin{center}
\begin{tabular}{ccccccccc}
\multicolumn{9}{c}{{\bf Table 1:} {\it - continued.}} \\
\hline
\hline
IUE Image & Observation & Exposure  & Observatory & $F_\nu^a$ & 
$\sigma_{F_\nu}$ & $\alpha_\nu^b$ & $\sigma_{\alpha_\nu}$ & $\chi^2_\nu$ \\
& Midpoint (UT) & Time & (Goddard & (mJy) & (mJy) & & & \\
& (day of May 94) & (min) & or Vilspa) & & & & &\\
\hline 
 LWP 28203 & 20.41673 & 25 & G &  20.28 &  0.38 &  1.07 &  0.14 &  0.68 \\  
 LWP 28204 & 20.48396 & 25 & G &  20.07 &  0.37 &  0.83 &  0.14 &  0.55 \\  
 LWP 28205 & 20.55041 & 25 & G &  20.07 &  0.37 &  0.89 &  0.14 &  0.66 \\  
 LWP 28206 & 20.61706 & 25 & G &  20.62 &  0.41 &  0.68 &  0.16 &  0.72 \\  
 LWP 28207 & 20.68032 & 25 & G &  22.14 &  0.77 &  0.94 &  0.34 &  0.63 \\  
 LWP 28208 & 20.74675 & 16 & G &  20.73 &  0.83 &  0.12 &  0.39 &  0.65 \\  
 LWP 28209 & 20.81692 & 15 & G &  22.24 &  0.43 &  0.43 &  0.15 &  0.64 \\  
 LWP 28210 & 20.94994 & 25 & G &  23.09 &  0.41 &  1.01 &  0.13 &  0.59 \\  
 LWP 28211 & 21.01677 & 25 & V &  22.71 &  0.40 &  0.83 &  0.13 &  0.65 \\  
 LWP 28212 & 21.08339 & 25 & V &  22.82 &  0.40 &  0.89 &  0.13 &  0.80 \\  
 LWP 28213 & 21.15039 & 25 & V &  23.02 &  0.41 &  1.03 &  0.13 &  0.68 \\  
 LWP 28214 & 21.21723 & 25 & V &  22.82 &  0.41 &  0.88 &  0.13 &  0.66 \\  
 LWP 28215 & 21.28325 & 25 & V &  22.70 &  0.41 &  0.81 &  0.13 &  0.66 \\  
 LWP 28216 & 21.35069 & 25 & G &  22.91 &  0.41 &  0.96 &  0.13 &  0.68 \\  
 LWP 28217 & 21.41593 & 25 & G &  22.44 &  0.40 &  0.77 &  0.13 &  0.53 \\  
 LWP 28218 & 21.48215 & 25 & G &  22.27 &  0.40 &  0.87 &  0.13 &  0.74 \\  
 LWP 28219 & 21.54946 & 25 & G &  22.25 &  0.40 &  0.71 &  0.13 &  0.56 \\  
 LWP 28220 & 21.61635 & 25 & G &  21.59 &  0.41 &  0.25 &  0.15 &  0.91 \\  
 LWP 28221 & 21.67990 & 25 & G &  22.72 &  0.75 &  0.91 &  0.32 &  0.64 \\  
 LWP 28222$^c$&21.74510&16 & G &  21.35 &  0.96 &  1.16 &  0.45 &  0.61 \\  
\hline  
\end{tabular}
\end{center}
\begin{center}
\begin{tabular}{ccccccccc}
\multicolumn{9}{c}{{\bf Table 1:} {\it - continued.}} \\
\hline
\hline
IUE Image & Observation & Exposure  & Observatory & $F_\nu^a$ & 
$\sigma_{F_\nu}$ & $\alpha_\nu^b$ & $\sigma_{\alpha_\nu}$ & $\chi^2_\nu$ \\
& Midpoint (UT) & Time & (Goddard & (mJy) & (mJy) & & & \\
& (day of May 94) & (min) & or Vilspa) & & & & &\\
\hline 
 LWP 28223 & 21.81732 & 13 & G &  19.85 &  0.40 &  0.45 &  0.16 &  0.63 \\  
 LWP 28224 & 21.95083 & 25 & G &  18.65 &  0.35 &  1.12 &  0.15 &  0.58 \\  
 LWP 28225 & 22.01694 & 25 & V &  18.46 &  0.35 &  1.14 &  0.15 &  0.64 \\  
 LWP 28226 & 22.08223 & 25 & V &  18.07 &  0.36 &  1.30 &  0.16 &  0.70 \\  
 LWP 28227 & 22.34867 & 25 & V &  17.06 &  0.34 &  1.09 &  0.16 &  0.61 \\  
 LWP 28228 & 22.41740 & 25 & G &  17.44 &  0.35 &  1.24 &  0.16 &  0.73 \\  
 LWP 28229 & 22.48184 & 25 & G &  17.45 &  0.35 &  1.52 &  0.17 &  0.65 \\  
 LWP 28230 & 22.54847 & 25 & G &  17.18 &  0.34 &  1.17 &  0.16 &  0.65 \\  
 LWP 28231 & 22.61508 & 25 & G &  17.92 &  0.38 &  0.73 &  0.18 &  0.80 \\  
 LWP 28232 & 22.67991 & 25 & G &  17.64 &  0.56 &  0.80 &  0.31 &  0.79 \\  
 LWP 28233 & 22.74640 & 20 & G &  17.19 &  0.65 &  0.96 &  0.38 &  0.69 \\  
 LWP 28234 & 22.81508 & 18 & G &  17.28 &  0.34 &  0.73 &  0.15 &  0.74 \\  
 LWP 28235 & 22.94809 & 25 & G &  17.63 &  0.34 &  1.04 &  0.15 &  0.63 \\  
 LWP 28236 & 23.01624 & 25 & V &  18.14 &  0.34 &  0.87 &  0.15 &  0.69 \\  
 LWP 28237 & 23.06579 & 25 & V &  18.45 &  0.35 &  1.30 &  0.15 &  0.73 \\  
 LWP 28238 & 23.14927 & 25 & V &  18.25 &  0.35 &  1.14 &  0.15 &  0.73 \\  
 LWP 28239 & 23.21620 & 25 & V &  18.34 &  0.35 &  1.09 &  0.15 &  0.55 \\  
 LWP 28240 & 23.28145 & 25 & V &  17.83 &  0.35 &  0.92 &  0.15 &  0.69 \\  
 LWP 28241 & 23.34773 & 25 & G &  18.28 &  0.35 &  1.20 &  0.15 &  0.72 \\  
 LWP 28242 & 23.41432 & 25 & G &  18.16 &  0.35 &  1.24 &  0.16 &  0.67 \\  
\hline  
\end{tabular}
\end{center}
\begin{center}
\begin{tabular}{ccccccccc}
\multicolumn{9}{c}{{\bf Table 1:} {\it - continued.}} \\
\hline
\hline
IUE Image & Observation & Exposure  & Observatory & $F_\nu^a$ & 
$\sigma_{F_\nu}$ & $\alpha_\nu^b$ & $\sigma_{\alpha_\nu}$ & $\chi^2_\nu$ \\
& Midpoint (UT) & Time & (Goddard & (mJy) & (mJy) & & & \\
& (day of May 94) & (min) & or Vilspa) & & & & &\\
\hline 
 LWP 28243 & 23.48093 & 25 & G &  18.39 &  0.35 &  1.26 &  0.15 &  0.71 \\  
 LWP 28244 & 23.54752 & 25 & G &  18.39 &  0.36 &  1.28 &  0.16 &  0.63 \\  
 LWP 28245 & 23.61417 & 25 & G &  18.15 &  0.37 &  0.74 &  0.17 &  0.77 \\  
 LWP 28246 & 23.68069 & 25 & G &  17.59 &  0.54 &  0.52 &  0.29 &  0.69 \\  
 LWP 28247 & 23.74660 & 25 & G &  18.83 &  0.58 &  0.85 &  0.29 &  0.54 \\  
 LWP 28248 & 23.81392 & 21 & G &  18.95 &  0.36 &  1.12 &  0.15 &  0.54 \\  
 LWP 28249 & 23.94704 & 25 & G &  18.69 &  0.36 &  1.18 &  0.15 &  0.64 \\  
 LWP 28250 & 24.01432 & 25 & V &  18.64 &  0.35 &  1.17 &  0.15 &  0.69 \\  
 LWP 28251 & 24.08010 & 25 & V &  18.85 &  0.36 &  1.28 &  0.15 &  0.74 \\  
 LWP 28252 & 24.14703 & 25 & V &  18.66 &  0.35 &  1.06 &  0.15 &  0.66 \\  
 LWP 28253 & 24.21411 & 25 & V &  19.57 &  0.37 &  1.35 &  0.15 &  0.79 \\  
 LWP 28254 & 24.28127 & 25 & V &  19.09 &  0.36 &  1.14 &  0.15 &  0.84 \\  
 LWP 28256 & 24.67891 & 25 & V &  20.00 &  0.37 &  0.81 &  0.14 &  0.80 \\  
 LWP 28257 & 24.74557 & 25 & G &  20.43 &  0.38 &  0.80 &  0.14 &  0.77 \\  
 LWP 28258 & 24.81200 & 25 & G &  20.66 &  0.38 &  1.30 &  0.14 &  0.68 \\  
 LWP 28259 & 24.94266 & 25 & G &  19.59 &  0.42 &  1.16 &  0.18 &  0.57 \\  
 LWP 28262 & 25.67900 & 17 & G &  20.34 &  0.39 &  0.52 &  0.15 &  0.66 \\  
 LWP 28263 & 25.74547 & 25 & G &  20.86 &  0.40 &  0.65 &  0.15 &  0.62 \\  
\hline  
\multicolumn{9}{l}{Note.- The number of degrees of freedom for the fits is
typically 410 for the SWP and}\\
\multicolumn{9}{l}{ 264 for the LWP.}\\
\multicolumn{9}{l}{$^a$ At 1400 \AA\ or 2800 \AA\ for SWP or LWP spectra,
respectively.}\\
\multicolumn{9}{l}{$^b$ Fitted ranges are 1230-1950 \AA\ and 2100-2800 \AA\
for SWP and LWP spectra, respectively.}\\
\multicolumn{9}{l}{$^c$ Underexposed spectrum.}\\
\end{tabular}
\end{center}

\newpage
%
%
\begin{center}
\begin{tabular}{cccccccc}
\multicolumn{8}{c}{{\bf Table 2:} Power-Law Fit Parameters of Merged SWP-LWP 
Spectra}\\
\hline
\hline
\multicolumn{2}{c}{Spectral Pair} & Observation & $F_{2000}$ & 
$\sigma_{F_{2000}}$ & $\alpha_C$ & $\sigma_{\alpha_C}$ & $\chi^2_\nu$ \\
\multicolumn{2}{c}{Image Numbers} & Midpoint (UT) & (mJy) & (mJy) & & & \\
SWP & LWP & (Day of May 94) & & & & &\\
\hline 
 50779 & 28143 & 16.16939 &  13.32 &   0.18 &   0.83 &   0.04 &  0.88 \\  
 50780 & 28144 & 16.23701 &  13.43 &   0.18 &   0.84 &   0.04 &  0.95 \\  
 50781 & 28145 & 16.30453 &  13.60 &   0.18 &   0.88 &   0.04 &  0.90 \\  
 50782 & 28146 & 16.36971 &  13.51 &   0.18 &   0.87 &   0.04 &  0.85 \\  
 50783 & 28147 & 16.43697 &  13.49 &   0.18 &   0.85 &   0.04 &  1.06 \\  
 50784 & 28148 & 16.50320 &  13.38 &   0.18 &   0.84 &   0.04 &  1.03 \\  
 50785 & 28149 & 16.56976 &  13.54 &   0.19 &   0.87 &   0.04 &  1.00 \\  
 50786 & 28150 & 16.63905 &  14.01 &   0.20 &   0.89 &   0.05 &  0.85 \\  
 50787 & 28151 & 16.70290 &  13.75 &   0.22 &   0.96 &   0.07 &  0.49 \\  
 50788 & 28152 & 16.76824 &  13.92 &   0.22 &   0.95 &   0.07 &  0.74 \\  
 50789 & 28153 & 16.83202 &  14.07 &   0.19 &   0.86 &   0.04 &  0.81 \\  
 50790 & 28154 & 16.97156 &  13.62 &   0.18 &   0.82 &   0.04 &  1.16 \\  
 50791 & 28155 & 17.03536 &  13.95 &   0.19 &   0.89 &   0.04 &  1.15 \\  
 50792 & 28156 & 17.10154 &  13.79 &   0.18 &   0.84 &   0.04 &  1.12 \\  
 50793 & 28157 & 17.16822 &  13.75 &   0.18 &   0.86 &   0.04 &  1.12 \\  
 50794 & 28158 & 17.23492 &  13.84 &   0.19 &   0.84 &   0.04 &  1.03 \\  
 50795 & 28159 & 17.30233 &  14.11 &   0.19 &   0.86 &   0.04 &  0.98 \\  
 50796 & 28160 & 17.36928 &  14.01 &   0.19 &   0.86 &   0.04 &  1.07 \\  
 50797 & 28161 & 17.43554 &  14.05 &   0.19 &   0.85 &   0.04 &  1.21 \\  
 50798 & 28162 & 17.50215 &  14.10 &   0.19 &   0.88 &   0.04 &  1.15 \\  
 50799 & 28163 & 17.56883 &  14.25 &   0.19 &   0.87 &   0.04 &  1.00 \\  
 50800 & 28164 & 17.63511 &  14.03 &   0.20 &   0.96 &   0.04 &  0.81 \\  
\hline 
\end{tabular}
\end{center}
\pagebreak
\begin{center}
\begin{tabular}{cccccccc}
\multicolumn{8}{c}{{\bf Table 2:} {\it - continued}}\\
\hline
\hline
\multicolumn{2}{c}{Spectral Pair} & Observation & $F_{2000}$ & 
$\sigma_{F_{2000}}$ & $\alpha_C$ & $\sigma_{\alpha_C}$ & $\chi^2_\nu$ \\
\multicolumn{2}{c}{Image Numbers} & Midpoint (UT) & (mJy) & (mJy) & & & \\
SWP & LWP & (Day of May 94) & & & & &\\
\hline 
 50801 & 28165 & 17.69974 &  13.85 &   0.23 &   0.90 &   0.08 &  0.70 \\ 
 50802 & 28166 & 17.76314 &  14.42 &   0.25 &   1.00 &   0.09 &  0.70 \\  
 50803 & 28167 & 17.83215 &  14.11 &   0.19 &   0.98 &   0.04 &  0.98 \\  
 50804 & 28168 & 17.96642 &  13.56 &   0.18 &   0.96 &   0.04 &  0.85 \\  
 50805 & 28169 & 18.03042 &  13.70 &   0.18 &   0.89 &   0.04 &  1.20 \\  
 50806 & 28170 & 18.09446 &  13.34 &   0.18 &   0.91 &   0.04 &  1.03 \\  
 50808 & 28171 & 18.36811 &  13.25 &   0.18 &   0.90 &   0.04 &  1.02 \\  
 50809 & 28172 & 18.43370 &  13.41 &   0.18 &   0.95 &   0.04 &  1.14 \\  
 50810 & 28173 & 18.50090 &  13.13 &   0.18 &   0.90 &   0.04 &  1.16 \\  
 50811 & 28174 & 18.56951 &  13.07 &   0.18 &   0.90 &   0.04 &  1.00 \\  
 50812 & 28175 & 18.63421 &  13.05 &   0.18 &   0.92 &   0.04 &  0.83 \\  
 50813 & 28176 & 18.69677 &  13.14 &   0.22 &   1.02 &   0.08 &  0.70 \\  
 50814 & 28177 & 18.76174 &  13.09 &   0.22 &   0.88 &   0.09 &  0.68 \\  
 50816 & 28183 & 19.03461 &  12.81 &   0.17 &   0.89 &   0.04 &  1.03 \\  
 50817 & 28184 & 19.10021 &  12.74 &   0.17 &   0.89 &   0.04 &  1.01 \\  
 50818 & 28185 & 19.16821 &  12.79 &   0.17 &   0.85 &   0.04 &  1.12 \\  
 50819 & 28186 & 19.23239 &  12.96 &   0.18 &   0.85 &   0.04 &  1.03 \\  
 50821 & 28188 & 19.36642 &  12.92 &   0.18 &   0.86 &   0.04 &  0.93 \\  
 50822 & 28189 & 19.43350 &  12.90 &   0.18 &   0.88 &   0.04 &  0.86 \\  
 50823 & 28190 & 19.49983 &  12.98 &   0.18 &   0.81 &   0.04 &  1.05 \\  
 50824 & 28191 & 19.56608 &  13.20 &   0.18 &   0.84 &   0.04 &  1.13 \\  
 50825 & 28192 & 19.63337 &  13.04 &   0.18 &   0.86 &   0.05 &  0.65 \\  
\hline 
\end{tabular}
\end{center}
\pagebreak
\begin{center}
\begin{tabular}{cccccccc}
\multicolumn{8}{c}{{\bf Table 2:} {\it - continued}}\\
\hline
\hline
\multicolumn{2}{c}{Spectral Pair} & Observation & $F_{2000}$ & 
$\sigma_{F_{2000}}$ & $\alpha_C$ & $\sigma_{\alpha_C}$ & $\chi^2_\nu$ \\
\multicolumn{2}{c}{Image Numbers} & Midpoint (UT) & (mJy) & (mJy) & & & \\
SWP & LWP & (Day of May 94) & & & & &\\
\hline 
 50826 & 28193 & 19.69594 &  12.84 &   0.22 &   0.93 &   0.09 &  0.58 \\  
 50827 & 28194 & 19.76089 &  13.21 &   0.23 &   0.96 &   0.09 &  0.73 \\  
 50828 & 28195 & 19.83112 &  13.37 &   0.18 &   0.95 &   0.04 &  0.90 \\  
 50829 & 28196 & 19.96575 &  13.60 &   0.18 &   0.84 &   0.04 &  1.01 \\  
 50830 & 28197 & 20.03319 &  14.08 &   0.19 &   0.83 &   0.04 &  0.86 \\  
 50831 & 28198 & 20.09990 &  14.49 &   0.19 &   0.83 &   0.04 &  0.99 \\  
 50832 & 28199 & 20.16596 &  14.58 &   0.20 &   0.79 &   0.04 &  1.01 \\  
 50833 & 28200 & 20.23311 &  15.03 &   0.20 &   0.83 &   0.04 &  0.93 \\  
 50834 & 28201 & 20.29966 &  14.88 &   0.20 &   0.83 &   0.04 &  1.05 \\  
 50835 & 28202 & 20.36594 &  15.13 &   0.20 &   0.84 &   0.04 &  1.14 \\  
 50836 & 28203 & 20.43272 &  15.33 &   0.20 &   0.91 &   0.04 &  1.29 \\  
 50837 & 28204 & 20.49974 &  15.37 &   0.20 &   0.87 &   0.04 &  0.94 \\  
 50838 & 28205 & 20.56612 &  15.48 &   0.21 &   0.86 &   0.04 &  1.26 \\  
 50839 & 28206 & 20.63278 &  15.26 &   0.21 &   0.99 &   0.05 &  0.72 \\  
 50841 & 28208 & 20.76070 &  16.20 &   0.28 &   0.92 &   0.09 &  0.65 \\  
 50842 & 28209 & 20.82994 &  17.44 &   0.23 &   0.93 &   0.04 &  1.02 \\  
 50843 & 28210 & 20.96575 &  17.31 &   0.23 &   0.91 &   0.03 &  1.18 \\  
 50844 & 28211 & 21.03256 &  17.49 &   0.23 &   0.89 &   0.03 &  1.06 \\  
 50845 & 28212 & 21.09926 &  17.41 &   0.23 &   0.88 &   0.03 &  1.28 \\  
 50846 & 28213 & 21.16615 &  17.44 &   0.23 &   0.84 &   0.03 &  0.90 \\  
 50847 & 28214 & 21.23246 &  17.46 &   0.23 &   0.84 &   0.04 &  1.06 \\  
 50848 & 28215 & 21.29944 &  17.52 &   0.23 &   0.86 &   0.03 &  1.06 \\  
\hline 
\end{tabular}
\end{center}
\pagebreak
\begin{center}
\begin{tabular}{cccccccc}
\multicolumn{8}{c}{{\bf Table 2:} {\it - continued}}\\
\hline
\hline
\multicolumn{2}{c}{Spectral Pair} & Observation & $F_{2000}$ & 
$\sigma_{F_{2000}}$ & $\alpha_C$ & $\sigma_{\alpha_C}$ & $\chi^2_\nu$ \\
\multicolumn{2}{c}{Image Numbers} & Midpoint (UT) & (mJy) & (mJy) & & & \\
SWP & LWP & (Day of May 94) & & & & &\\
\hline 
 50849 & 28216 & 21.36610 &  17.28 &   0.23 &   0.90 &   0.03 &  1.09 \\  
 50850 & 28217 & 21.43163 &  17.10 &   0.23 &   0.90 &   0.04 &  1.16 \\  
 50851 & 28218 & 21.49824 &  17.04 &   0.22 &   0.84 &   0.03 &  1.16 \\  
 50852 & 28219 & 21.56520 &  16.99 &   0.23 &   0.86 &   0.04 &  0.80 \\  
 50853 & 28220 & 21.63216 &  16.16 &   0.23 &   1.07 &   0.05 &  0.93 \\  
 50856 & 28223 & 21.82905 &  14.75 &   0.20 &   1.08 &   0.04 &  1.17 \\  
 50857 & 28224 & 21.96654 &  13.73 &   0.19 &   0.95 &   0.04 &  1.06 \\  
 50858 & 28225 & 22.03217 &  13.54 &   0.18 &   0.95 &   0.04 &  1.01 \\  
 50859 & 28226 & 22.09781 &  12.92 &   0.18 &   1.00 &   0.04 &  0.96 \\  
 50861 & 28227 & 22.36525 &  12.75 &   0.17 &   0.94 &   0.04 &  0.94 \\  
 50862 & 28228 & 22.43277 &  12.69 &   0.17 &   0.96 &   0.04 &  1.03 \\  
 50863 & 28229 & 22.49748 &  12.70 &   0.17 &   0.91 &   0.04 &  1.27 \\  
 50864 & 28230 & 22.56410 &  12.81 &   0.17 &   0.88 &   0.04 &  0.86 \\  
 50865 & 28231 & 22.63074 &  12.81 &   0.18 &   1.05 &   0.05 &  0.72 \\  
 50866 & 28232 & 22.69263 &  12.81 &   0.22 &   1.02 &   0.09 &  0.67 \\  
 50867 & 28233 & 22.76174 &  12.60 &   0.20 &   0.94 &   0.08 &  0.69 \\  
 50868 & 28234 & 22.82696 &  12.55 &   0.17 &   1.05 &   0.04 &  1.01 \\  
 50869 & 28235 & 22.96410 &  13.32 &   0.18 &   0.88 &   0.04 &  1.19 \\  
 50870 & 28236 & 23.03028 &  13.68 &   0.18 &   0.93 &   0.04 &  1.10 \\  
 50871 & 28237 & 23.08910 &  13.64 &   0.18 &   0.90 &   0.04 &  1.47 \\  
 50872 & 28238 & 23.16539 &  13.63 &   0.18 &   0.90 &   0.04 &  1.07 \\  
 50873 & 28239 & 23.23104 &  13.61 &   0.18 &   0.93 &   0.04 &  0.98 \\  
\hline 
\end{tabular}
\end{center}
\pagebreak
\begin{center}
\begin{tabular}{cccccccc}
\multicolumn{8}{c}{{\bf Table 2:} {\it - continued}}\\
\hline
\hline
\multicolumn{2}{c}{Spectral Pair} & Observation & $F_{2000}$ & 
$\sigma_{F_{2000}}$ & $\alpha_C$ & $\sigma_{\alpha_C}$ & $\chi^2_\nu$ \\
\multicolumn{2}{c}{Image Numbers} & Midpoint (UT) & (mJy) & (mJy) & & & \\
SWP & LWP & (Day of May 94) & & & & &\\
\hline 
 50874 & 28240 & 23.29742 &  12.95 &   0.18 &   1.04 &   0.04 &  1.18 \\  
 50875 & 28241 & 23.36335 &  12.95 &   0.17 &   1.05 &   0.04 &  0.98 \\  
 50876 & 28242 & 23.43009 &  13.23 &   0.18 &   0.97 &   0.04 &  1.08 \\  
 50877 & 28243 & 23.49656 &  13.46 &   0.18 &   0.91 &   0.04 &  0.99 \\  
 50878 & 28244 & 23.56297 &  13.58 &   0.18 &   0.89 &   0.04 &  0.99 \\  
 50879 & 28245 & 23.62984 &  13.08 &   0.19 &   1.03 &   0.05 &  0.86 \\  
 50880 & 28246 & 23.69373 &  12.73 &   0.21 &   1.11 &   0.08 &  0.67 \\  
 50881 & 28247 & 23.76311 &  13.20 &   0.20 &   1.06 &   0.07 &  0.90 \\  
 50882 & 28248 & 23.82520 &  13.05 &   0.18 &   1.16 &   0.04 &  0.97 \\  
 50883 & 28249 & 23.96293 &  13.88 &   0.18 &   0.91 &   0.04 &  1.14 \\  
 50884 & 28250 & 24.03092 &  13.03 &   0.17 &   1.05 &   0.04 &  0.97 \\  
 50885 & 28251 & 24.09589 &  13.91 &   0.18 &   0.90 &   0.04 &  1.14 \\  
 50886 & 28252 & 24.16287 &  14.12 &   0.19 &   0.87 &   0.04 &  1.10 \\  
 50887 & 28253 & 24.23064 &  14.17 &   0.19 &   0.92 &   0.04 &  1.12 \\  
 50889 & 28256 & 24.69450 &  15.47 &   0.21 &   0.86 &   0.04 &  1.28 \\  
 50890 & 28257 & 24.76122 &  15.38 &   0.20 &   0.96 &   0.03 &  1.11 \\  
 50891 & 28258 & 24.82330 &  14.85 &   0.20 &   0.96 &   0.04 &  1.04 \\  
 50894 & 28262 & 25.69474 &  15.53 &   0.21 &   0.93 &   0.04 &  0.98 \\  
 50895 & 28263 & 25.75818 &  15.74 &   0.21 &   0.97 &   0.04 &  0.85 \\  
\hline  
\multicolumn{8}{l}{Note.- The number of degrees of freedom for the fits is
typically 625.}
\end{tabular}
\end{center}
 
\newpage
%
%
\begin{center}
\begin{tabular}{cccc}
\multicolumn{4}{c}{{\bf Table 3:} Comparison between SWET and TOMSIPS
Extractions}\\
\hline
\hline
Range & \multicolumn{3}{c}{Average Spectral Indices} \\
              & 1991, SWET$^a$ & 1991, TOMSIPS & 1994, TOMSIPS \\
\hline
1230-1950 \AA & 0.80$\pm$0.06$^b$ (98$^c$) & 0.91$\pm$0.06 (98) & 
1.02$\pm$0.08 (115) \\
2100-2800 \AA &      --       & 0.94$\pm$0.13 (97) & 0.95$\pm$0.28 (117) \\
2100-3100 \AA & 0.83$\pm$0.12 (97) & 0.83$\pm$0.10 (97) &      --       \\
1230-2800 \AA &      --       & 0.83$\pm$0.04 (99) & 0.91$\pm$0.07 (107) \\
1230-3100 \AA & 0.63$\pm$0.04 (99) & 0.81$\pm$0.04 (99) &      --       \\
\hline 
\multicolumn{4}{l}{$^a$ Results from U93.}\\
\multicolumn{4}{l}{$^b$ Standard deviation with respect to the mean. Typical
errors on individual}\\
\multicolumn{4}{l}{~~ spectral indices of the November 1991 campaign were
0.05 for SWP, 0.1}\\
\multicolumn{4}{l}{~~ for LWP and 0.02 for merged spectra.}\\
\multicolumn{4}{l}{$^c$ Number of spectra.}\\

\end{tabular}
\end{center}



\newpage

\cl{\bf Figure Captions}

\bigskip

{\bf Fig. 1} -- Typical spectra from the May 1994 IUE campaign not corrected 
for reddening (upper panels). In both cases the power-law fitting curve 
$F_\lambda \propto \lambda^{-\beta}$ is 
shown as a solid line, with indices {\it a)} $\beta$ = 0.92$\pm$0.03; {\it b)} 
$\beta$ = 0.76$\pm$0.13. The lower panels represent the intrinsic 
error distributions of the spectral fluxes.

{\bf Fig. 2} -- Histograms of dereddened SWP (solid line) and LWP 
(dashed line) spectral indices for {\it a)} $A_V$ = 0.1 mag; {\it b)} 
$A_V$ = 0.4 mag.

{\bf Fig. 3} -- Dereddened light curves at 1400 \AA\ (filled circles) and 2800 
\AA\ (open circles): {\it a)} full observing period (the circled points 
correspond to underexposed spectra, see text); {\it b)} expanded view of the 
initial portion. In the second plot, the light curves are normalized to their
respective averages, calculated after excluding the flux points taken during 
the first day of monitoring. Variability is detected on time scales comparable
to the exposure times (up to a factor $\sim$2.2 flux change at 2800 \AA\ in 
1.5 hr), and more rapid variations are probably present but unresolved.

{\bf Fig. 4} -- Auto-correlation function of the 1400 \AA\ flux (filled 
circles) and the 2800 \AA\ flux (open circles) computed with the DCF of 
Edelson \& Krolik (1988).

{\bf Fig. 5} -- Cross-correlation function between the 1400 \AA\ and 2800 \AA\
light curves computed with the DCF of Edelson \& Krolik (1988).

{\bf Fig. 6} -- Spectral indices for the dereddened flux 
distributions in the {\it a)} 1230-1950 \AA\ band; {\it b)} 2100-2800 \AA\ 
band; {\it c)} 1230-2800 \AA\ band. The circled points correspond to 
underexposed spectra. The horizontal solid lines represent the average
energy indices in each band.

{\bf Fig. 7} -- Cross-correlation function between the flux at 1400 \AA\
and the SWP spectral index computed with the DCF of Edelson \& Krolik (1988). 
The spectrum flattens $\sim$1 day before the flux increases.

{\bf Fig. 8} -- Comparison of the light curves at 1400 \AA\ (filled) 
and at 2800 \AA\ (open) obtained during the present IUE campaign (circles) 
and during the intensive monitoring period in November 1991 (squares). 
Day 1 in the temporal scale corresponds to 10 November for the 1991 data and 
to 15 May for the 1994 data. For both epochs, the light curves have been 
normalized to the average SWP and LWP fluxes in 1991. The flux level in 1991 
was $\sim$20\% brighter than in 1994. The character of the 
variability is different at the two epochs: recurrent $\sim$20\%
variations detected in 1991 are not seen in the 1994 data, which exhibit
an extremely rapid flux doubling at the beginning of the light curve and a 
big central flare of $\sim$35\% amplitude.


\begin{references}

\ref{Allen, R. G., Smith, P. S., Angel, J. R. P., Miller, B. W., Anderson,
S. F., \& Margon, B. 1993, ApJ, 403, 610}

\ref{Ayres, T. R. 1993, PASP, 105, 538}

\ref{Ayres, T. R., et al. 1995, ApJS, 96, 223}

\ref{Blandford, R. D. 1990, in Active Galactic Nuclei, Eds. T. J.-L.
Courvoisier and M. Mayor (Berlin: Springer Verlag), p. 168}

\ref{Bonnell, J. T., Vestrand, W. T., \& Stacy, J. G. 1994, ApJ, 420, 545} 

\ref{Bregman, J. N. 1990, A\&AR, 2, 125}

\ref{Bruhweiler, F. C., Boggess, A., Norman, D. J., Grady, C. A., Urry, C. M.,
\& Kondo, Y. 1993, ApJ, 409, 199}

\ref{Caplinger, J. 1995, NASA IUE Newsletter No. 55, 17}

\ref{Cappi, M., Comastri, A., Molendi, S., Palumbo, G. G. C., Della Ceca, R.,
\& Maccacaro, T. 1994, MNRAS, 271, 438}

\ref{Cardelli, J. A., Clayton, G. C., \& Mathis, J. S. 1989, ApJ, 345, 245}

\ref{Carini, M., \& Miller, H. R. 1992, ApJ, 385, 146}

\ref{Celotti, A., Maraschi, L., \& Treves, A. 1991, ApJ, 377, 403}

\ref{Crenshaw, M. D., Bruegman, O. W., \& Norman, D. J. 1990, PASP, 102, 463}

\ref{Edelson, R. A., \& Krolik, J. H. 1988, ApJ, 333, 646}

\ref{Edelson, R. A., et al. 1991, ApJ, 372, L9}

\ref{Edelson, R. A. 1992, ApJ, 401, 516}

\ref{Edelson, R. A., Pike, G. F., Saken, J. M., Kinney, A., \& Shull, J. M.
1992, ApJS, 83, 1}

\ref{Edelson, R. A., et al. 1995, ApJ, 438, 120}

\ref{Fabian, A. C. 1979, Proc. Roy. Soc., 366, 449}

\ref{Falomo, R., Pesce, J. E., \& Treves 1993, ApJ, 411, L63}

\ref{Feigelson, E. D., et al. 1986, ApJ, 302, 337}

\ref{Georganopoulos, M., \& Marscher, A. P. 1996, in Proc. of the Workshop 
on Blazar Continuum Variability, Miami, February 1996, p. 262}

\ref{Ghisellini, G., Maraschi, L., \& Treves, A. 1985, A\&A, 146, 204}

\ref{Ghisellini, G., et al. 1997, A\&A, accepted}

\ref{Hutter, D. J., \& Mufson, S. L. 1986, ApJ, 301, 50}

\ref{Imhoff, C. 1996, IUE NASA Newsletter, 56, 108}

\ref{Kii, T., et al. 1996, in preparation}

\ref{Kinney, A. L., Bohlin, R. C., \& Neill, J. D. 1991a, PASP, 103, 694}

\ref{Kinney, A. L., Bohlin, R. C., Blades, J. C., \&  York, D. G. 1991b, ApJS, 
75, 645}

\ref{K\"onigl, A. 1981, ApJ, 243, 700}

\ref{Koratkar, A. P., Pesce, J. E., Urry, C. M., \& Pian, E. 1996, ApJ,
 submitted}

\ref{Lockman, F. J., \& Savage, B. D. 1995, ApJS, 97, 1}

\ref{Macomb, D. J., et al. 1995, ApJ, 449, L99}

\ref{Maraschi, L., Tagliaferri, G., Tanzi, E. G., \& Treves, A. 
1986, ApJ, 304, 637}

\ref{Marscher, A. P. 1980, ApJ, 235, 386}

\ref{Marshall, H., et al. 1996, in preparation}

\ref{Mattox, J. R., Wagner, S. J., Malkan, M., McGlynn, T. A., Schachter, J. 
F., Grove, J. E., Johnson, W. N., \& Kurfess, J. D. 1997, ApJ, 476, 692}

\ref{Miller, H. R., \& Noble, J. C. 1996, in Blazar Continuum Variability,
ASP Conf. Ser. Vol. 110, eds H. R. Miller, J. R. Webb, and J. C. Noble, p. 17}

\ref{Morini, M., et al. 1986, ApJ, 306, L71} 

\ref{Nichols, J. S., \& Linsky, J. L. 1996, AJ, 111, 517}


\ref{Paltani, S., Courvoisier, T. J.-L., Bratschi, P., \& Blecha, A. 1996, 
Proc. of the Workshop on Blazar Continuum Variability, Miami, February 1996, 
p. 36}

\ref{Pesce, J. E., et al. 1997, ApJ, in press}

\ref{Pian, E., \& Treves, A. 1993, ApJ, 416, 130}

\ref{Pian, E., Treves, A., Webb, J., Kazanas, D., Maraschi, L., McCollum,
B., Shrader, C., \& Wamsteker, W. 1996,  Proceedings of the OJ-94 Project 
Meeting, Tuorla Observatory Reports, No. 176, Ed. L. O. Takalo, p. 26}

\ref{Rieke, G. H., \& Lebofsky, M. J. 1985, ApJ, 288, 618}

\ref{Seaton, M. J. 1979, MNRAS, 187, 73p}

\ref{Sembay, S., Warwick, R. S., Urry, C. M., Sokoloski, J., George, I. M.,
Makino, F., \& Ohashi, T. 1993, ApJ, 404, 112}

\ref{Shrader, C. R., et al. 1994, AJ, 107, 904}

\ref{Shull, J. M., \& Van Steenberg, M. E. 1985, ApJ, 294, 599}

\ref{Sillanp\"a\"a, A. 1996, Proc. of the Workshop on Blazar Continuum 
Variability, Miami, February 1996, p. 74}
 
\ref{Smith, P. S. 1996, Proc. of the Workshop on Blazar Continuum 
Variability, Miami, February 1996, p. 135}

\ref{Stark, A. A., Gammie, C. F., Wilson, R. W., Bally, J., Linke, R. A., 
Heiles, C., \& Hurwitz, M. 1992, ApJS, 79, 77}

\ref{Treves, A., et al. 1989, ApJ, 341, 733}

\ref{Treves, A., \& Girardi, E. 1991, in Variability of Active Galaxies, 
eds. W. J. Duschl, S. J. Wagner, and M. Camenzind, Lecture Notes in
Physics, 377 (Springer Verlag: Berlin), p. 175}
 
\ref{Urry, C. M., Kondo, Y., Hackney, K. R. H., \& Hackney, R. L. 1988,
ApJ, 330, 791}

\ref{Urry, C. M., \& Reichert, G. A. 1988, IUE NASA Newsletter, 34, 95}

\ref{Urry, C. M., et al. 1993, ApJ, 411, 614 (U93)}

\ref{Urry, C. M., et al. 1997, ApJ, in press}

\ref{Vestrand, W. T., Stacy, J. G., \& Sreekumar, P. 1996, ApJ, 454, L93} 

\ref{Wagner, S. J., \& Witzel, A. 1994, in The Nature of Compact Objects in
Active Galactic Nuclei, Proc. of the 33rd Herstmonceux Conference, Eds.
A. Robinson and R. Terlevich, Cambridge University Press, p. 397}

\ref{Wagner, S. J., \& Witzel, A. 1995, ARAA, 33, 163}

\ref{Wandel, A., \& Urry, C. M. 1991, ApJ, 367, 78}

\end{references}
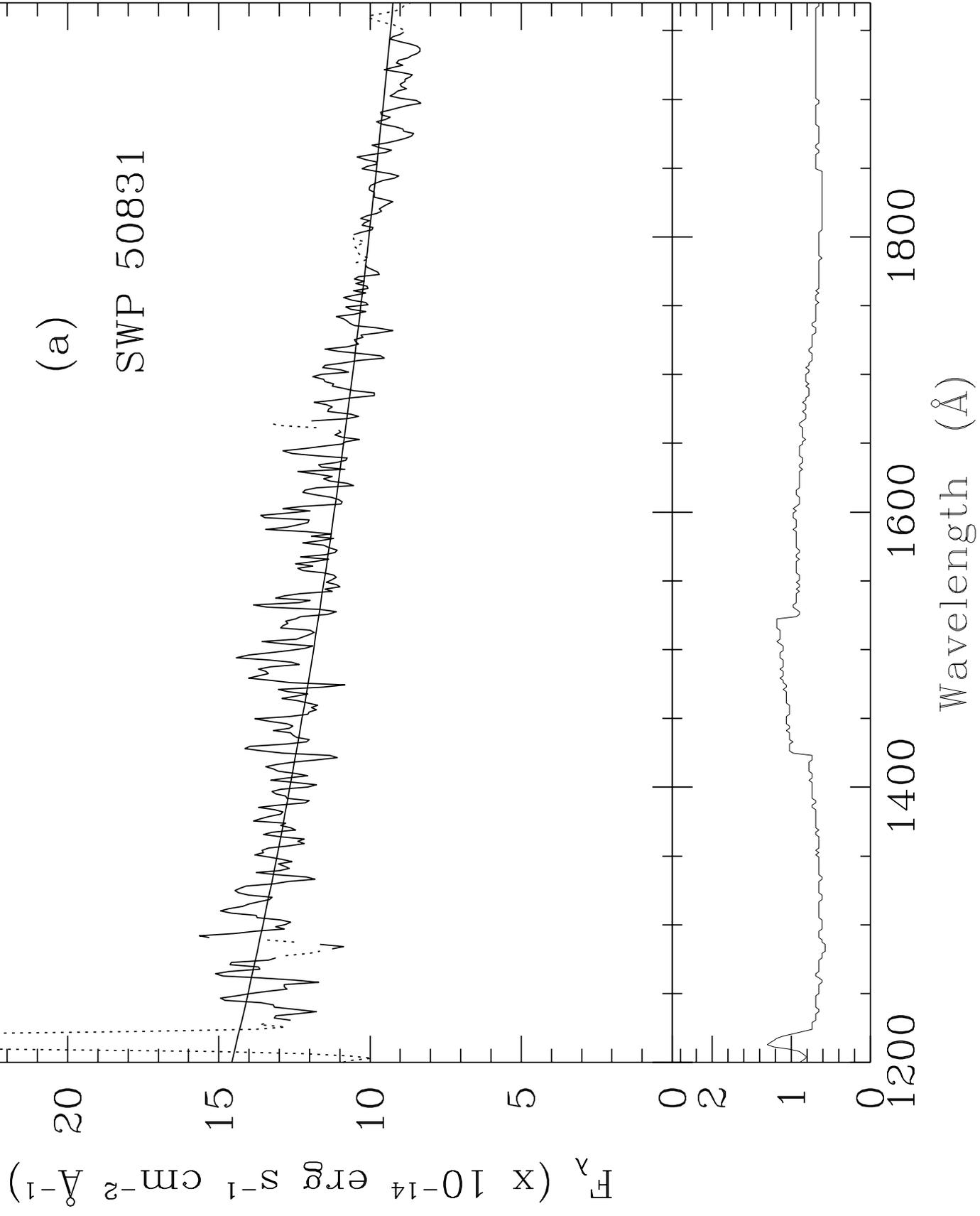
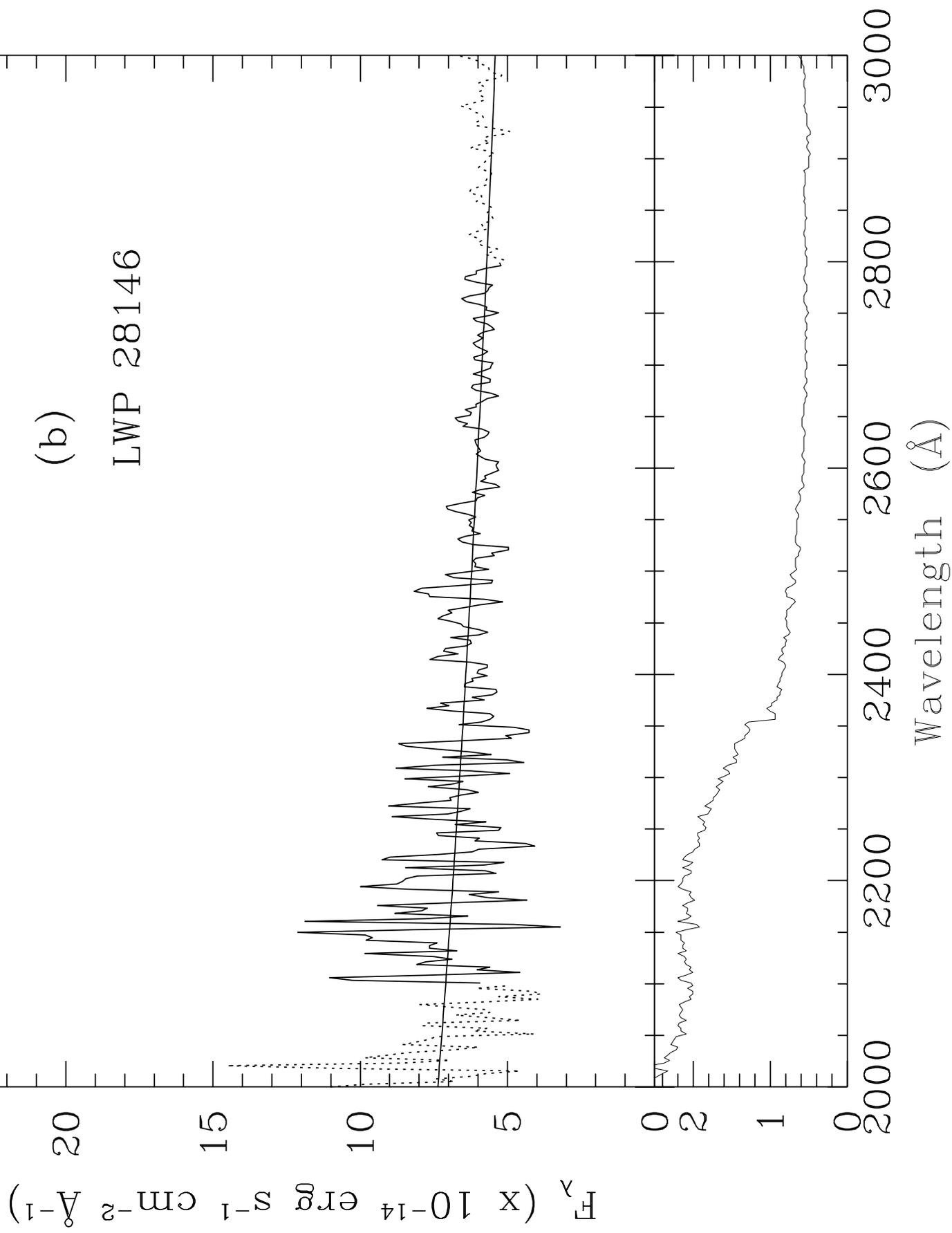
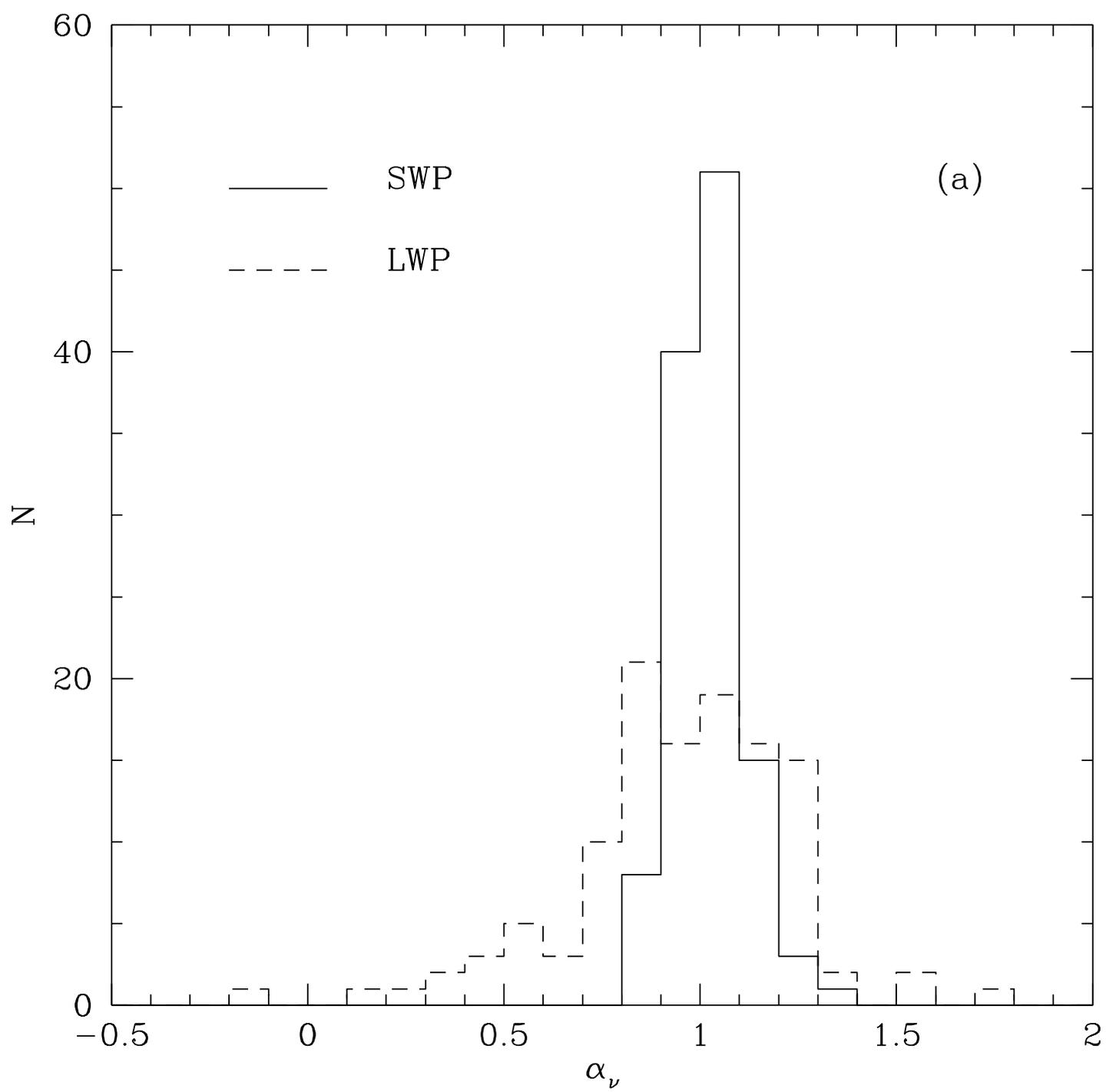
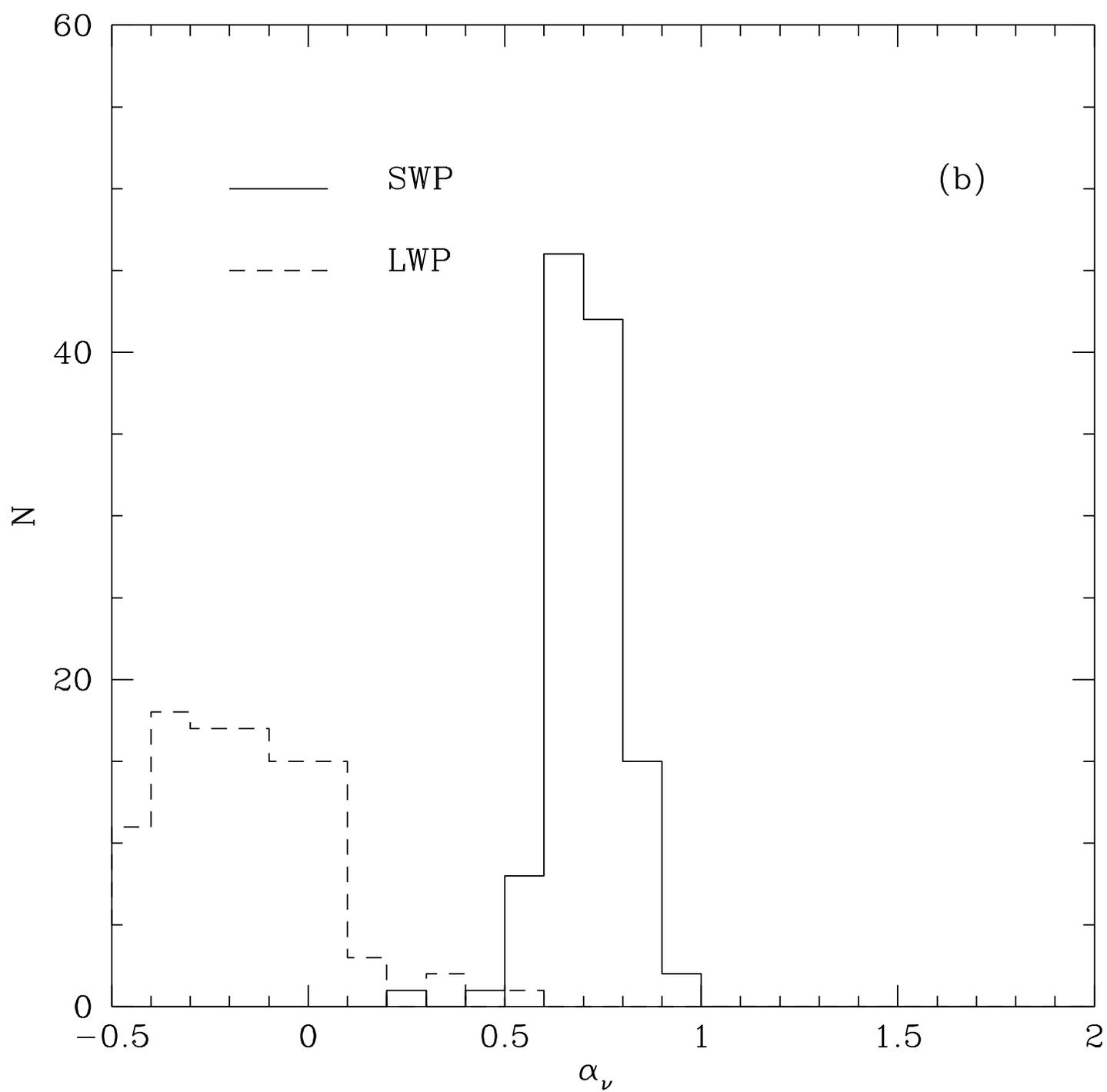
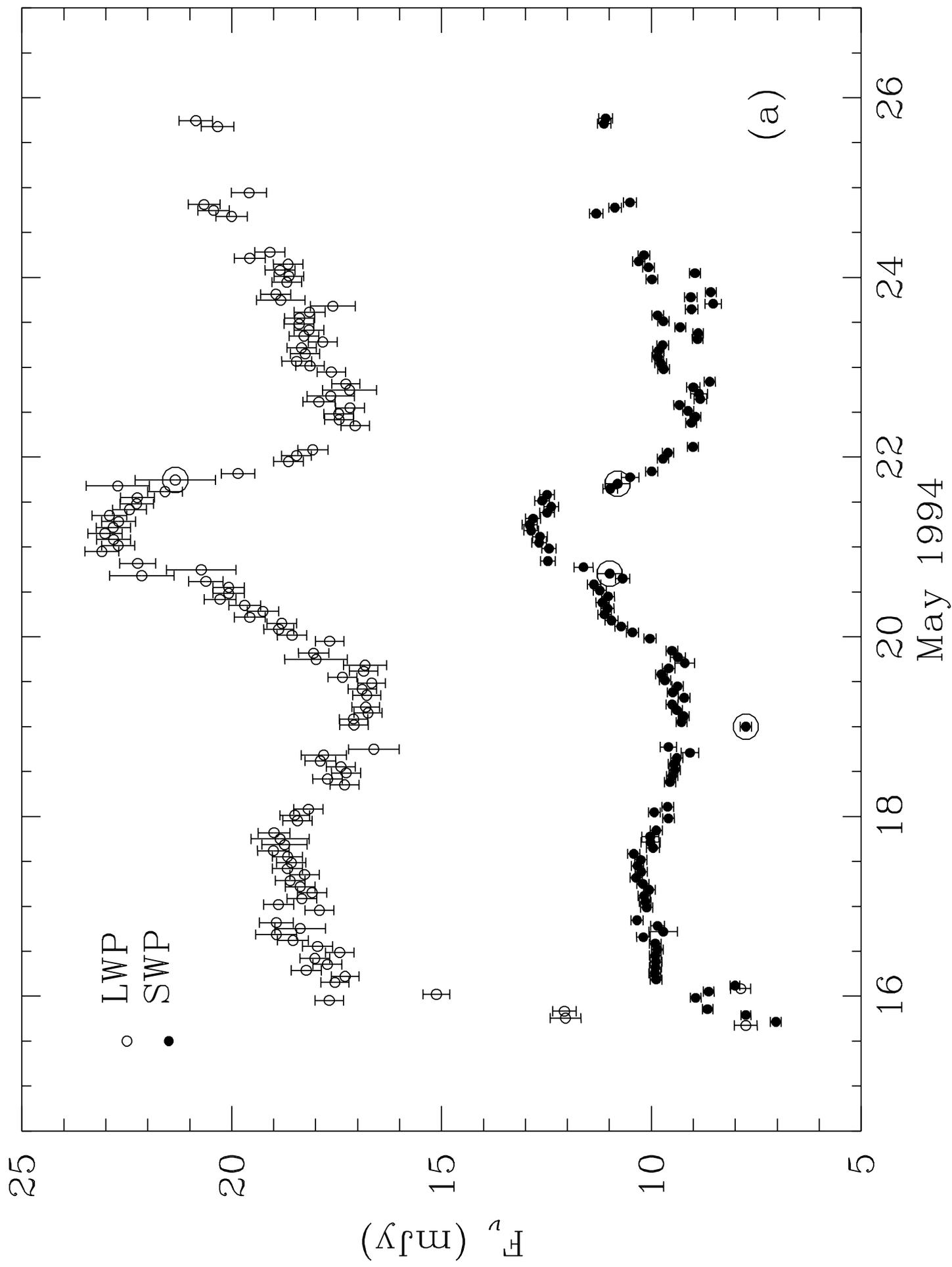
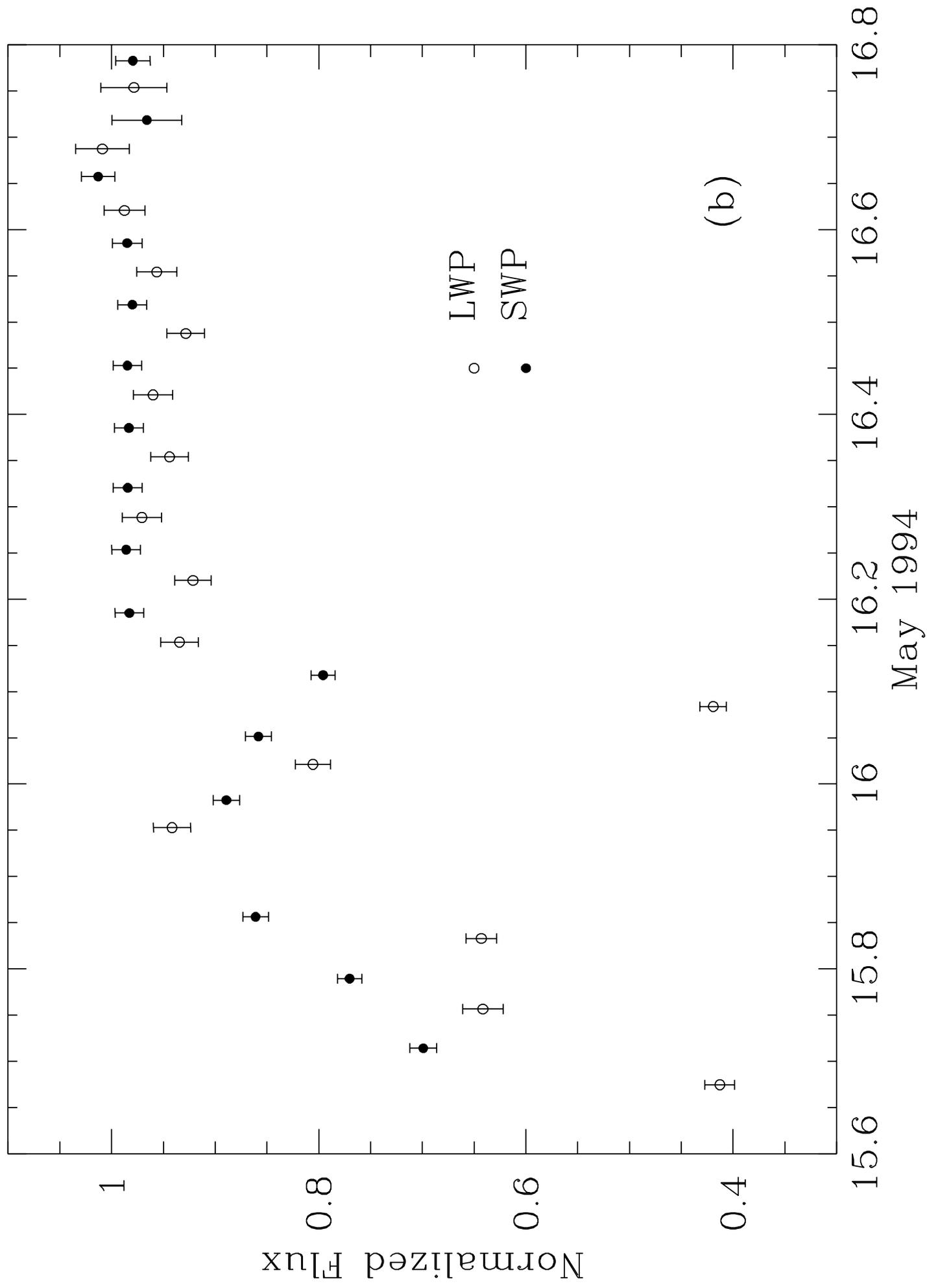
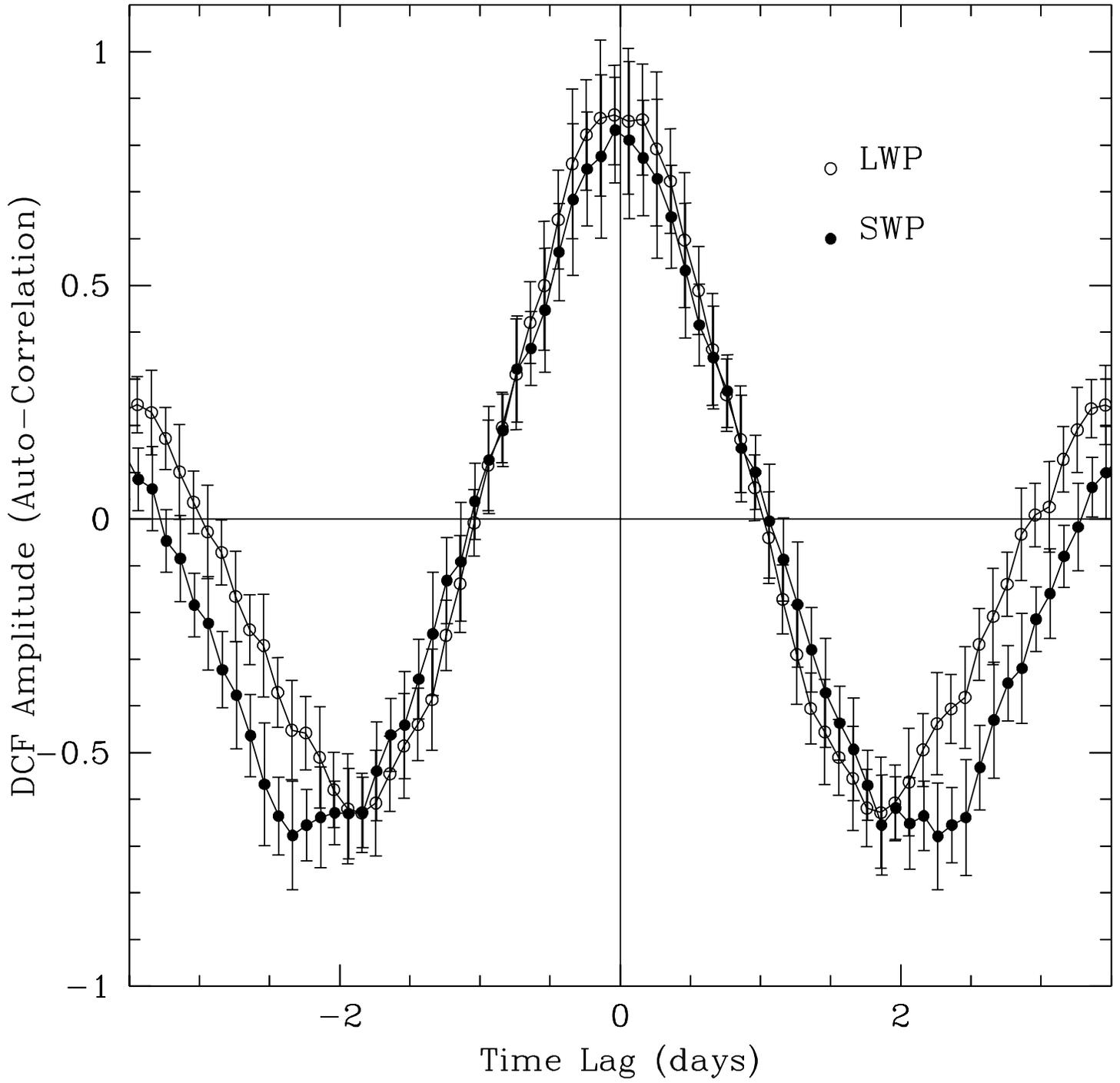
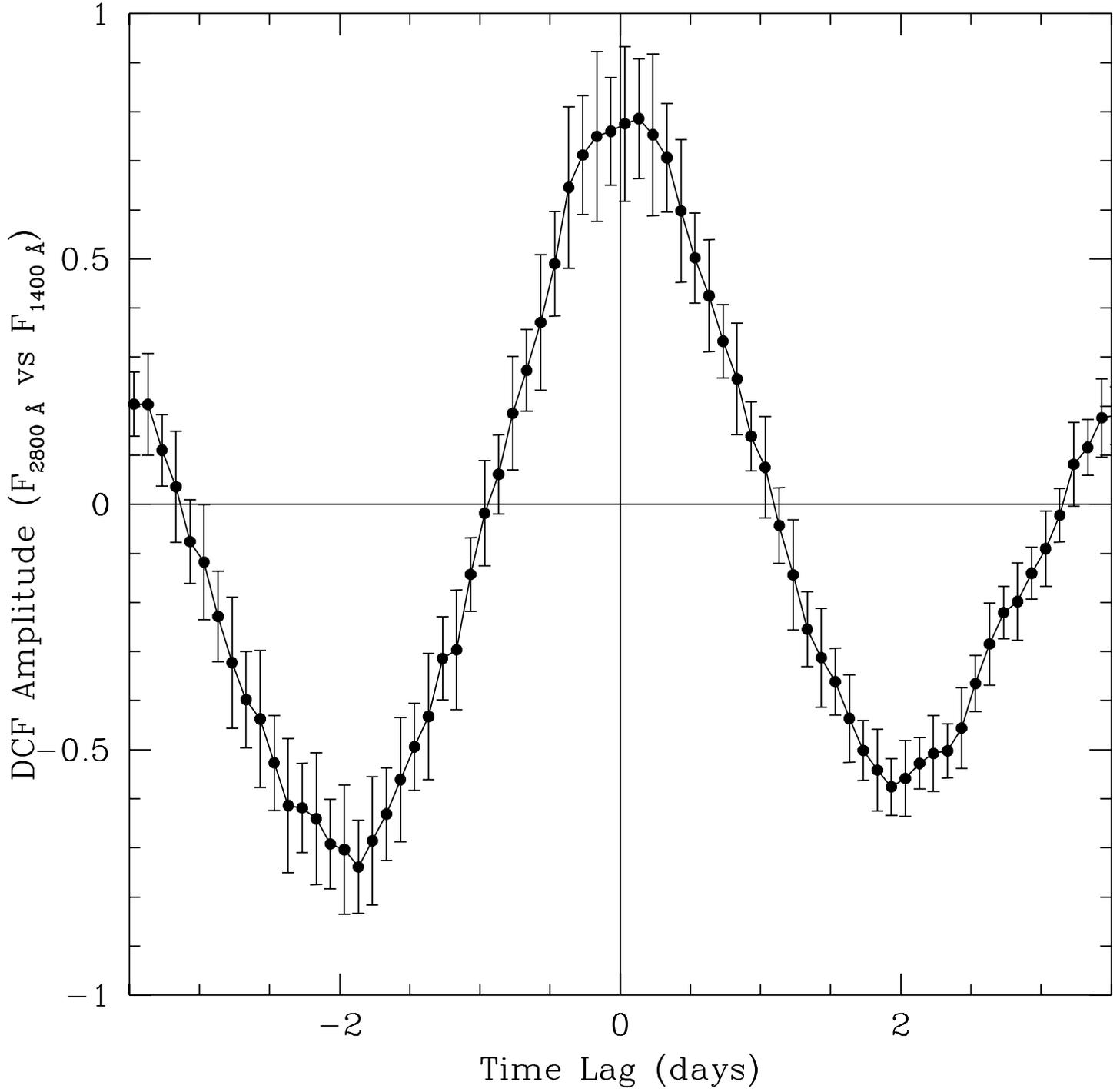
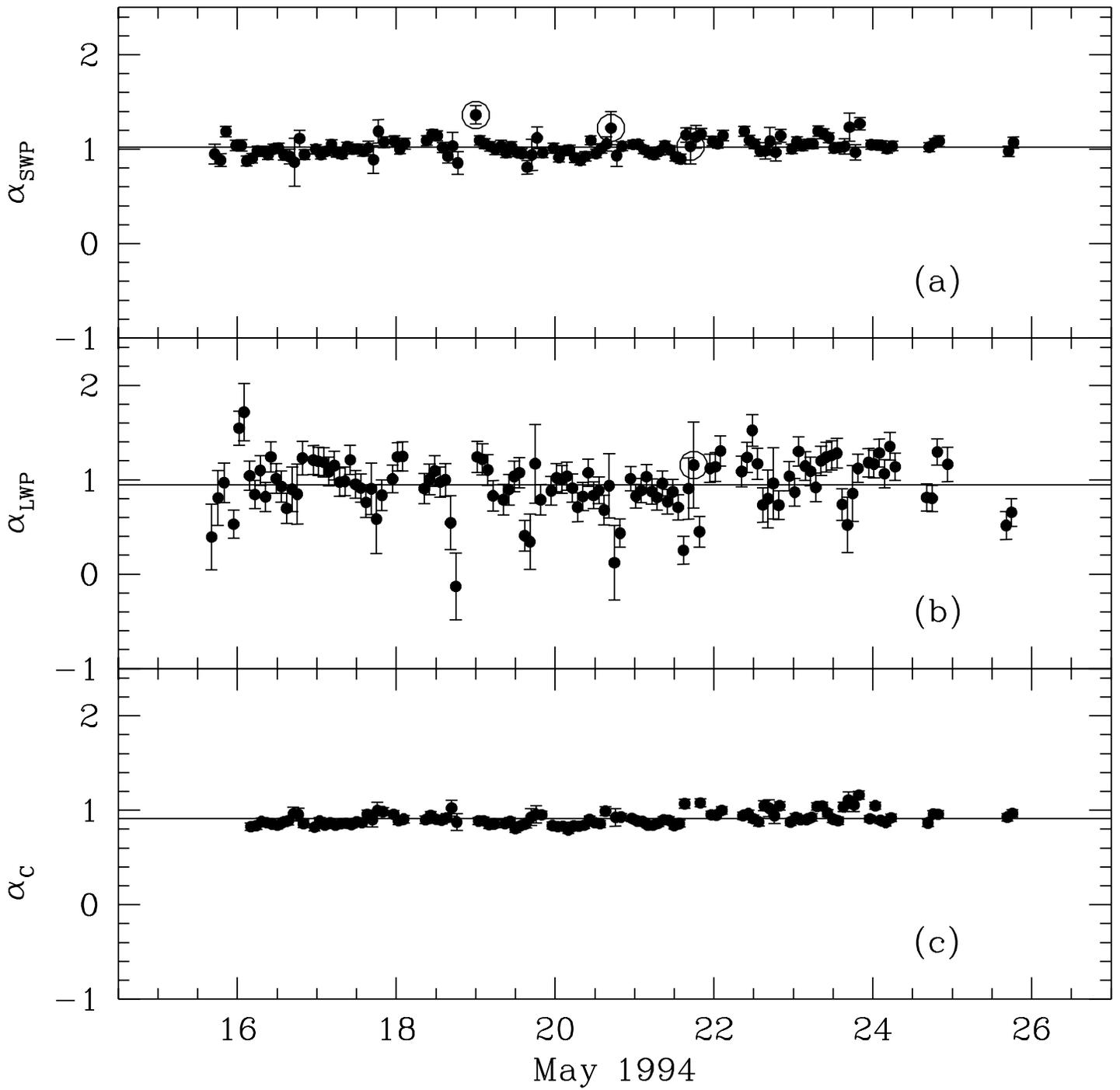
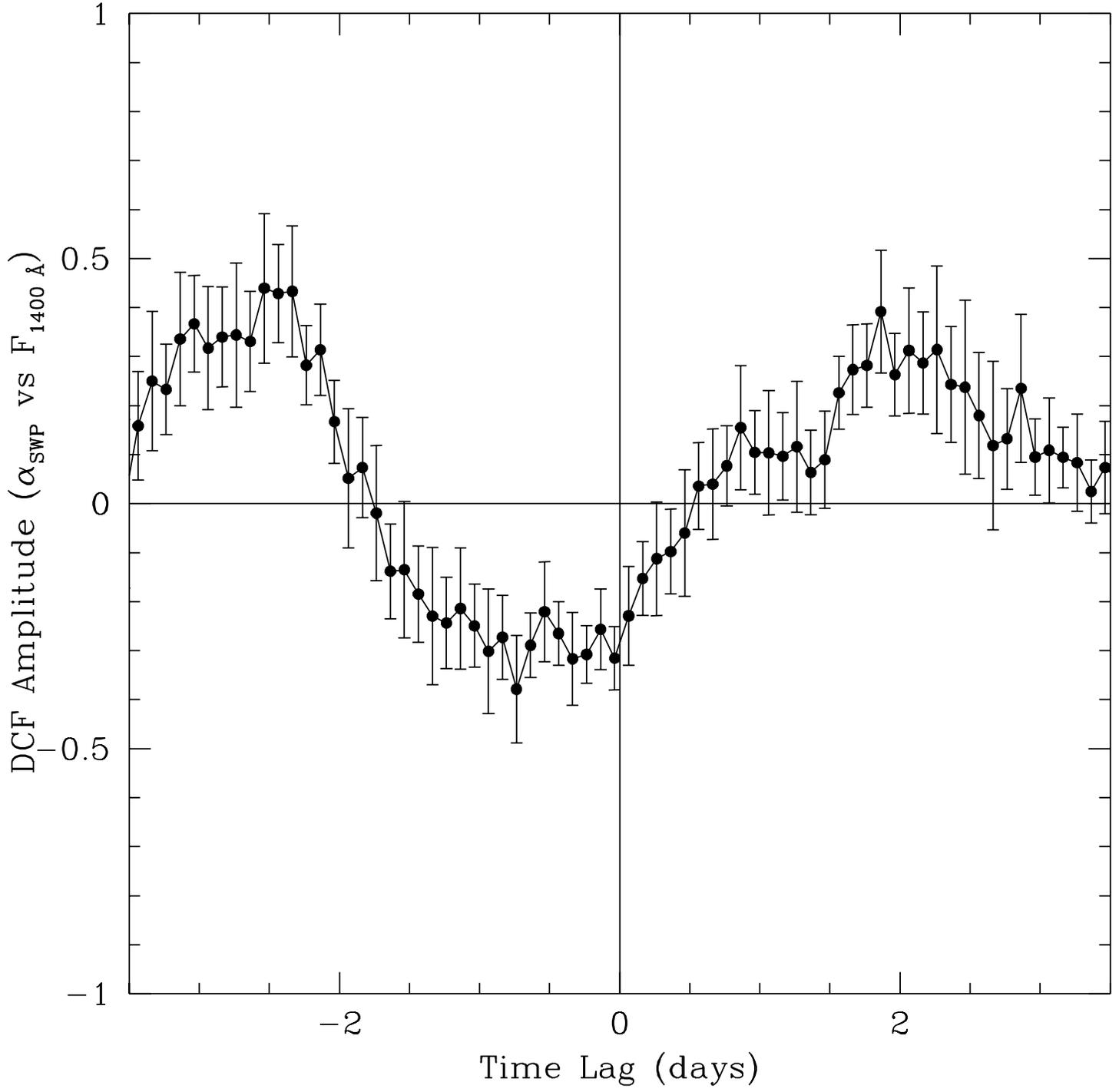
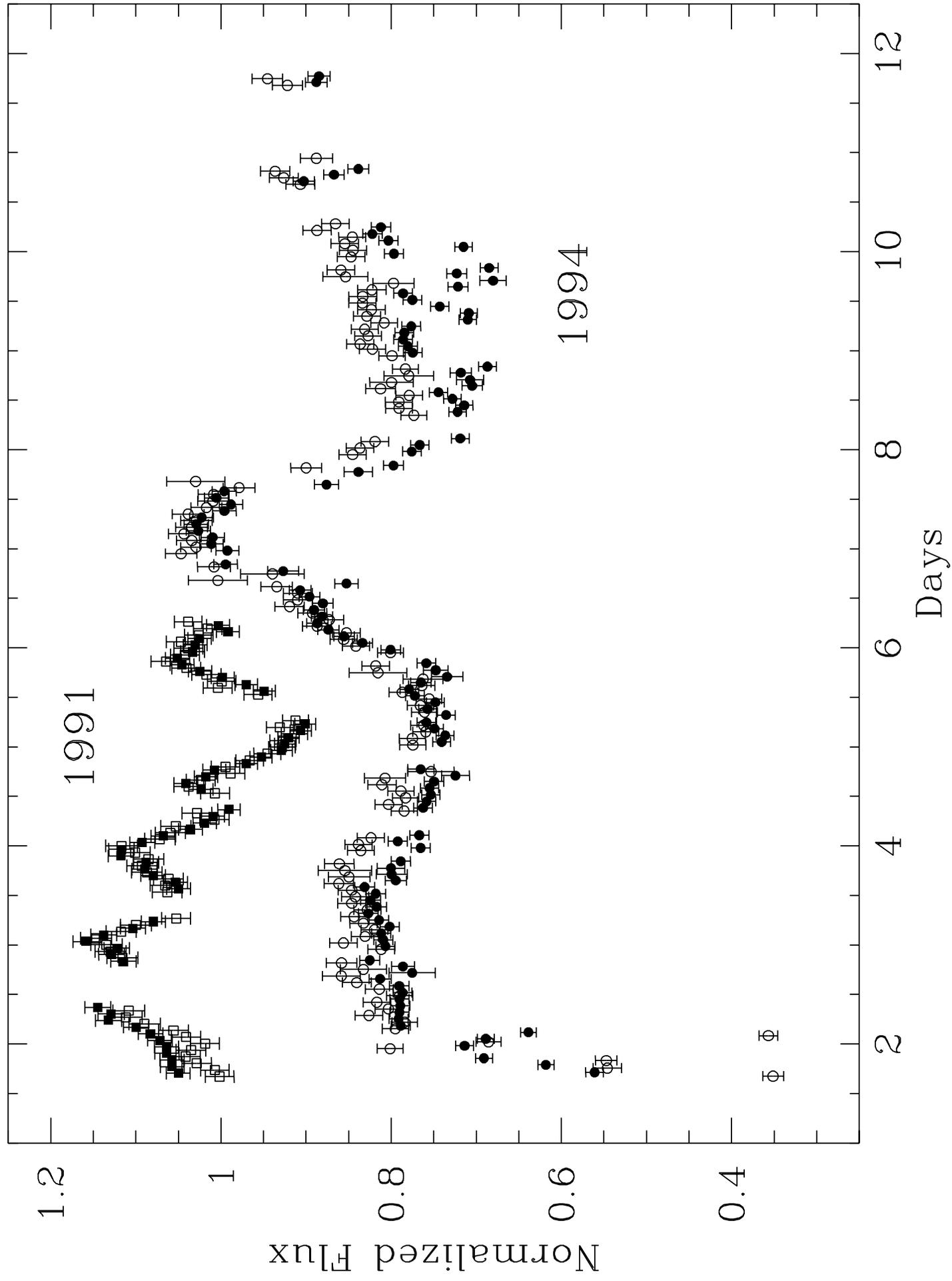
\end{document}